\documentclass[12pt,preprint]{aastex}

\begin{document}

\title{Infrared Spectroscopy of Symbiotic Stars. VII. \\
Binary Orbit and Long Secondary Period Variability of CH Cygni
}

\author{Kenneth H. Hinkle}
\affil{National Optical Astronomy Observatory, P.O. Box 26732,
Tucson, AZ USA}
\email{hinkle@noao.edu}

\author{Francis C. Fekel\altaffilmark{1}}
\affil{Center of Excellence in Information Systems, Tennessee
State University, Nashville, TN USA}
\email{fekel@evans.tsuniv.edu}

\author{Richard R. Joyce}
\affil{National Optical Astronomy Observatory, P.O. Box 26732, 
Tucson, AZ USA}
\email{joyce@noao.edu}

\altaffiltext{1}{Visiting Astronomer, Kitt Peak National Observatory,
National Optical Astronomy Observatory, operated by the Association of
Universities for Research in Astronomy, Inc., under cooperative
agreement with the National Science Foundation.}

\begin{abstract}

High-dispersion spectroscopic observations are used to refine orbital
elements for the symbiotic binary CH~Cyg.  The current radial
velocities, added to a previously published 13 year time series of
infrared velocities for the M giant in the CH Cyg symbiotic system,
more than double the length of the time series to 29 years.  The
two previously identified velocity periods are confirmed.  The long
period, revised to 15.6 $\pm$ 0.1 yr, is shown to result from a
binary orbit with a 0.7 M$_\odot$ white dwarf and 2 M$_\odot$ M
giant.  Mass transfer to the white dwarf is responsible for the
symbiotic classification.  CH Cyg is the longest period S-type
symbiotic known.  Similarities with the longer period D-type systems
are noted.  The 2.1 year period is shown to be on Wood's sequence
D, which contains stars identified as having long secondary periods
(LSP).  The cause of the LSP variation in CH Cyg and other stars
is unknown.  From our review of possible causes, we identify g-mode
non-radial pulsation as the leading mechanism for LSP variation in
CH Cyg.  If g-mode pulsation is the cause of the LSPs a radiative
region is required near the photosphere of pulsating AGB stars.

\end{abstract}


\section{INTRODUCTION}

Symbiotic stars are mass transfer binaries, containing a mass losing 
red giant and a mass accreting companion that is usually 
a white dwarf.  One system is known with a neutron star
\citep{hinkle_et_al_2006}, and systems containing main sequence
stars are possible but have proved elusive to identify.  
Mass transfer causes the symbiotics to exhibit highly complex light
variations and spectra.  From their characteristics at infrared 
wavelengths \citet{wa75} separate the symbiotics into two subclasses,
D for dusty-type and S for stellar-type systems.
CH~Cyg is the brightest of the S-type symbiotic stars.  

The current series of papers is concerned with infrared spectroscopy 
of the late-type star in each symbiotic system.
Because velocities of this component are usually well behaved, we 
employ these spectra to derive single-lined spectroscopic orbits.
The orbital elements provide insight into the period distribution, 
mass ratios, and orbital dynamics of these interacting systems.  These
results are ultimately of interest in evaluating the potential for 
catastrophic demise of the stellar components \citep{iben_2003}.

CH Cyg was the topic of our first paper on the infrared spectroscopy
of symbiotic stars \citep[][hereafter
Paper I]{hinkle_et_al_93}.  As a result of its brightness and
placement in the northern sky, CH Cyg is easily observed.  Following
the publication of Paper I, we continued to monitor CH Cyg with
various spectrographs and now have data spanning a period of nearly three 
decades.  In the intervening 16 years since the publication of Paper I
our knowledge of M giant variables as well as the CH Cyg system itself
has greatly increased.  We find, in agreement with many others who 
have worked on symbiotics \citep[see e.g.,][]{skopal_et_al_1996, 
crocker_et_al_2002, sokoloski_kenyon_2003b}, 
that CH Cyg is the most complex of the symbiotics that we have observed.

Based on velocity variations of the M III star in the CH Cyg system,
we concluded in Paper I that the CH~Cyg system is triple.
Two separate stable velocity variations, a long
period variation of $\sim$14.5 years and a short period
variation of $\sim$2.1 years, were found and are confirmed and refined 
in the present paper.  In the discussion section of Paper~I we proposed 
a model for this velocity behavior that placed the
symbiotic binary in the 2.1 year orbit.  Our model, based on the short
period mass function, had three underpinnings:

(1) The absence of any other known S-type symbiotic with a confirmed 
orbital period
longer than five years and a mean period distribution for S-type
symbiotic orbits of $\sim$2 years \citep{fekel_et_al_2007}.

(2) Stellar pulsation theory showing that the two year velocity
variation is too long to be the M giant fundamental pulsation
\citep{hughes_wood_1990}.

(3) Weak evidence at that time that the system had a high inclination.

Because this triple-star model has been controversial and new
information has accumulated, we decided to reexamine this
conclusion. In the present paper we review all the evidence and show
convincingly that the triple-star model  with the white dwarf as a
member of the short period system  is {\it incorrect}.  We now
conclude instead that the system is a binary with the mass accreting
white dwarf in the long period orbit.

Nonetheless, points (1) and (2) remain valid and are essential to
an understanding of CH~Cyg.  While the CH~Cyg symbiotic system has unique
properties, we believe the most interesting aspect of CH~Cyg is the
nature of the 2.1 year variation.  We show that CH~Cyg
qualifies as the most intensively investigated variable star with 
long secondary period (LSP) light and velocity variations.  The 
existence and uncertain nature of the LSP variations is highlighted 
by \citet{wood_2007}.

\section{OBSERVATIONS AND REDUCTIONS}

In 1992 we obtained two spectroscopic observations with the Fourier
transform spectrometer (FTS) at the coud\'e focus of the Kitt Peak
National Observatory (KPNO) 4 m telescope \citep{hetal78}.  These
observations are a continuation of the series of 71 FTS spectra
discussed in detail in Paper I.  The two observations
were obtained in the 2~$\mu$m region and have a resolving power of
$\sim$60,000.  Shortly after the two new spectra were obtained, the
FTS was decommissioned as a result of budget cuts at KPNO.

From 1995 through 2000 we collected 25 observations with the coud\'e
feed telescope and spectrograph system at KPNO.  The detector was
an infrared camera, NICMASS, developed at the University of
Massachusetts.  We obtained a 2 pixel resolving power of 44,000 at
a wavelength of 1.623 $\mu$m.  A more extensive description of the
experimental setup can be found in \citet{jetal98} and \citet{fetal00}.
The NICMASS equipment was sent to Mt. Stromlo Observatory to continue
the symbiotic orbit program in the southern hemisphere.  That equipment 
along with the 1.85 m telescope was destroyed in a devastating 
bush fire that burned over Mt. Stromlo in January 2003.

In 2000 three spectra were also acquired with the Phoenix cryogenic
echelle spectrograph at the f/15 Cassegrain focus of either the
KPNO 2.1 or 4 m telescopes.  A complete description of the spectrograph
can be found in \citet{hetal98}.  The observations were centered
at either 1.563 or 2.226 $\mu$m and have resolving powers of either
50,000 or 70,000.  An expanded discussion of the experimental setup
is in \citet{fetal00}.  Phoenix was deployed to the 
southern hemisphere in 2000.  

Finally, five spectrograms were obtained in 2007 and 2008 with the KPNO
coud\'e feed telescope, coud\'e spectrograph, and a CCD with enhanced 
red sensitivity, designated LB1A.  This 1980 x 800 pixel CCD was 
manufactured by Lawrence Berkeley National Laboratory and is 300~$\mu$m 
thick.  Although this thickness results in increased pixel contamination 
by cosmic-ray and background events, the chip was used because of its 
high quantum efficiency at far-red wavelengths.  Our spectrograms, 
centered near 1.005~$\mu$m, have a wavelength range of 420~\AA\ and a 
resolving power of $\sim$21,500.

Reduction and radial velocity measurement of the FTS spectra are
discussed in detail in Paper I.  For the NICMASS
and Phoenix data standard observing and reduction techniques were
used \citep{joyce_92}.  Wavelength calibration at the infrared wavelengths
of 1.563, 1.623, and 2.226~$\mu$m posed a challenge, because the
spectral coverage was far too small to include a sufficient number
of ThAr emission lines for a dispersion solution.  Thus, our approach
was to utilize the absorption lines of a K~III star.  Several sets
of lines were tried, including CO, Fe~I, and Ti~I.  These groups
all gave consistent results.  For the spectrograms acquired with
the LB1A CCD at 1.005 $\mu$m, we were able to use ThAr spectra for
wavelength calibration.  Telluric lines are present in the 1.005
and 2.226 $\mu$m wavelength regions.  These lines were removed from
our observations by ratioing the spectra to a hot star spectrum
observed on the same night.

For all of our spectra except those obtained with the FTS, the
absorption line radial velocities of CH~Cyg were measured with the
IRAF cross-correlation program FXCOR \citep{f93}.  Those velocities
were determined relative to the M-giant International Astronomical
Union velocity standards $\delta$~Oph or $\alpha$~Cet, which have
radial velocities of $-$19.1 and $-$25.3 km~s$^{-1}$, respectively
\citep{sbf90}.  The standards were observed multiple times during
the course of each night.  For the FTS spectra, which cover a
wavelength range that is $\sim$100 times greater than the other spectra, 
velocities were referenced to telluric lines.  

\section{ORBITAL SOLUTION}

Our 106 KPNO observations, which span nearly 30 years, are listed in 
Table~1, and the velocities are plotted in Figure~1 versus heliocentric 
Julian date.  That plot clearly shows two periodicities.  Thus, we 
have employed the general least squares program of \citet{d66} to
obtain a simultaneous orbital solution for the short- and long-period
velocity variations of CH~Cyg.  All KPNO velocities were given unit
weight after a comparison of the FTS velocities with those from the
Phoenix and coud\'e feed spectrographs showed that the RMS values
of the velocity data sets were nearly identical.  We then computed
several different orbital solutions.  The first used only the KPNO 
velocities, while, similar to a solution of Paper I, the second included
the velocities of \citet{detal74}, \citet{ym79}, and \citet{yy91}
plus our KPNO velocities and is designated the all-velocity
solution.  Weights adopted for the non-KPNO velocities are the same
as those assigned in Paper I.  As was done in that paper, we have also 
computed orbital solutions with the short period eccentricity fixed
at 0.0.  However, given the systematic residuals that result from
adopting such a sinusoidal fit and our discussion and conclusion 
about whether this short period velocity variation results 
from orbital motion or pulsation, we do not discuss the $e$ = 0.0 
solutions any further.

A comparison of the two remaining solutions shows that the long period 
of the KPNO-only solution is nearly 40 days longer
than that of the all-velocity solution, but that difference is only
0.7\% and is less than the period uncertainty.  Uncertainties of the 
orbital parameters in the all-velocity solution are somewhat smaller 
than those in the KPNO-only solution, as would be expected because of the 
longer baseline of data, but the parameter values of both solutions 
are within those uncertainties.  Thus, in Table~2 we have chosen to 
adopt the more homogeneous solution that includes only our KPNO 
velocities.  

For the individual observations Table~1 lists the heliocentric
Julian date, the observed total velocity, and the observed minus
calculated velocity residual ($O-C$) to the combined orbit. Also
computed and listed in the table are the long period orbital phase,
the long period velocity, which is equal to the total velocity minus
the computed short period velocity, the short period orbital phase,
and finally, the short period velocity, which is equal to the total
velocity minus the computed long period velocity.  Figure~2 presents
the computed velocity curve of the long period orbit compared with
the KPNO radial velocities, where zero phase is a time of periastron.
Each plotted velocity consists of the total observed velocity minus
its calculated short period velocity.  Figure~3 shows the computed
velocity curve of the short period orbit compared with the KPNO
radial velocities, where zero phase is a time of periastron.  Each
plotted velocity consists of the total observed velocity minus its
calculated long period velocity.

The radial velocity residuals, listed in column 3 of Table
1 ($``O-C''$), were analyzed for possible additional periods.  The program
PeriodoGRAM (PGRAM) was employed to fit sinusoids to the phased
data for periods between 50 and 500 days in steps of 0.05 days.
This period finding program works well even for non-sinusoidal
curves with eccentricities as large as $\sim$0.4.  Various subsets
of data were searched in an attempt to remove false periods.  Two
possible periods of $\sim$62 and $\sim$170 days appear in the
velocity residuals. However, the velocity semiamplitudes are very
small, at maximum 0.4 km s$^{-1}$.  Of particular interest, we find
no period above the noise in the 90-110 day period range, which
corresponds to a photometric period of CH Cyg.

\citet{ym79} first proposed an orbit for the M giant. Despite the use 
of very heterogeneous data with large uncertainties, they determined a 
probable period of 5750 days, very close to our value of 5689 days.  
Although their semiamplitude and eccentricity are rather different from
ours, other elements are similar.  We also compare our orbital solution 
to those in Paper~I.  In that earlier analysis the long period for the 
all-velocity solution was 5298 days, while that for the KPNO velocities 
alone was 5483 days.  Thus, the improved period of our adopted solution 
(Table~2) is nearly 400 days longer than the all-velocity solution in 
Paper~I and about 200 days longer than the KPNO-only solution given in 
that paper.  Comparing other elements of our adopted solution and the 
KPNO-only solution of Paper~I, the orbital elements are similar, but 
the uncertainties of the long period elements have been significantly 
reduced.

The ephemeris for conjunctions with the M giant in front of the
white dwarf, which corresponds to times of mid-eclipse, is
\begin{equation}
T_{conj}(HJD) = 2,446,353(\pm192) + 5689(\pm47)E,
\end{equation}
where $E$ is an integer number of cycles.  The above ephemeris for 
the eclipse of the white dwarf by
the M giant predicts mid-eclipse dates of HJD 2,440,664 for 1 cycle 
earlier and HJD 2,452,042 for 1 cycle later.  The uncertainty of
these predictions is about 240 days.

\section{BASIC SYSTEM PARAMETERS}

Paper I presented a broad introduction to CH~Cyg, but much material has
been published subsequent to that time. To understand the nature of 
the CH Cyg system, we first undertake an extensive review of the literature 
to determine basic parameters: variability, distance, luminosity, 
radius, rotation, mass, mass loss, and orbital inclination.  This results in a revised
picture of the system.

\subsection{Variability}

The photometric variability of CH Cyg was reviewed by
\citet{muciek_mikolajewski_1989} and \citet{mikolajewski_et_al_1990}.
The mass accreting component in the CH~Cyg system contributes to
the flux especially at blue wavelengths.  During periods of strong
activity, the M giant variability can be cloaked even in the optical
with light generated by the accretion process.  However, during quiescent
periods, the M giant can be detected into the blue.

CH Cyg had an extended period of quiescence from 1885 -- 1963.
During this interval CH Cyg was known only as a late-type semiregular
variable star.  A review of historic material on the variability
of CH Cyg by \citet{muciek_mikolajewski_1989}, obtained during the
period of quiescence, finds that the dominant period in the M giant
is $\sim$100 days.  During periods of symbiotic activity the 100
day period can be hard to detect
\citep{mikolajewski_et_al_1992,munari_et_al_1996}.  The visual
amplitude of the 100 day period is small, $\sim$0.1 magnitude
\citep{muciek_mikolajewski_1989}.  The exact period depends on the
dates sampled.  \citet{mikolajewski_et_al_1990} gave the dominant
period as 94 and 99$\pm$2 days while \citet{mikolajewski_et_al_1992}
found 102$\pm$3 days.  \citet{mikolajewski_et_al_1992} concluded
that the $\sim$100 day period is systematically increasing at a
rate of $\sim$5 days per century and suggested, following the work
of \citet{wood_zarro_1981}, that the CH Cyg M giant is currently
undergoing a helium-shell flash.  \citet{mikolajewski_et_al_1990}
and \citet{mikolajewski_et_al_1992} both noted that the $\sim$100
day period implies that the CH~Cyg giant is a first overtone pulsator.

In addition to the 100 day period, a period of $\sim$770 days is
also frequently reported.  \citet{mikolajewski_et_al_1992} provided
a detailed analysis of the 770 day photometric period in CH~Cyg.
The photometric properties of the 770 day variation show increasing
amplitude toward shorter wavelengths but not in a way consistent
with dust extinction.  In addition, the energy distribution in the
IR suggests changing spectral type (and temperature) during the
light cycle, but there are no spectral type changes in the blue.
While \citet{munari_et_al_1996} failed to detect this period,
\citet{skopal_et_al_2007} noted the presence of a $\sim$750 day
period with amplitude of nearly 1 mag in $V$ photometry.  The $\sim$750
day period is apparent in data illustrated in \citet{skopal_et_al_2007}.
The symbiotic was quiescent at the time of these observations.
There also is a suggestion of the $\sim$100 day period in this data,
but the spacing of the data is too coarse for the period to be
adequately sampled.

The simultaneous presence of both periods in CH~Cyg was reported
by \citet{payne-gaposhkin_1954}.  She listed CH~Cyg as one of a
family of pulsators with a LSP and estimated a typical long to short
period ratio of $\sim$9 for these objects.  A longer list of LSP
semiregular variables appears in \citet{houk_1963}.
\citet{wood_olivier_kawaler_2004} asserted that the LSP stars listed
in \citet{payne-gaposhkin_1954} and \citet{houk_1963} are those
with the largest amplitude secondary periods.  The \citet{houk_1963}
catalog contains only $\sim$1.5\% of the then known long period
variables.  More recently, the fraction of local long period variables
with long secondary periods is reported by \citet{percy_et_al_2004}
and \citet{wood_olivier_kawaler_2004} as $\sim$25-30\%.

In paper I we noted the close agreement between the 770 day photometric
period and the 750 day spectroscopic period and assumed both originate
from the same physical mechanism.  We make the same assumption in
this paper.  \citet{hinkle_et_al_2002}, \citet{olivier_wood_2003},
and \citet{wood_olivier_kawaler_2004} detected LSP variations
spectroscopically in a number of semiregular variables.  The
semiregular 100 day pulsation was not detected spectroscopically.
Scaling from typical ratios of velocity to photometric amplitude
for SRb variables \citep{lebzelter_et_al_2000}, the CH Cyg 0.1
magnitude visual amplitude implies a $\sim$0.5 km s$^{-1}$ velocity
amplitude.  This is near the detection limit for our observations
(\S3).

\subsubsection{Eclipses} 

When the accretion disk in the CH Cyg system is active, light from
the disk dominates the blue and violet portions of the
spectrum.  At such times the optical light curve is highly complex.
\citet{eyres_et_al_2002} present a summary of recent optical
photometry.  Because CH~Cyg has a complex light curve, evidence
for eclipses depends critically on collaborating evidence rather
than just sudden decreases in the $U$ band light.  The first claim of an
eclipse of the hot component by the red giant, occurring from 1985 May
to October, was reported by \citet{mikolajewski_et_al_1987}.  This was
largely based on timing, matching a $U$ band event to one more than
15 years earlier.  However, these decreases in light show considerable 
structure because of activity in the system.  
The dates of mid-eclipse are JD 2440585 and JD 2446270 with the eclipse
FWHM $\sim$200 days. The ingress is brief, perhaps 10 days.  
\citet{mikolajewski_et_al_1990} reported an 
ephemeris of JD$_{min}$ = 2,446,275($\pm$75) + 5700($\pm$75)E.

Subsequent to the claim of \citet{mikolajewski_et_al_1987} considerable
evidence has been published that supports the claim that the 1985 $U$ 
band decrease is an eclipse.  \citet{skopal_et_al_1996} summarized a 
list of five changes in the CH Cyg system that were observed during the
1985 eclipse.  These are (1) the disappearance of rapid optical flickering,
(2) the dominance of the M giant continuum in optical spectra at
mid-eclipse, (3) the transition of double peaked Balmer emission lines
to weak broad single peaked emission lines at mid-eclipse, (4) the
dominance of the red side of the double peaked emission in H$\alpha$
and H$\beta$ before mid-eclipse and the dominance of the blue
emission following mid-eclipse, and (5) the prominence of the nebular
lines [Ne III] 3869 \AA\ and [O III] 5007 \AA.  Items (1) and (2)
show that the accretor was eclipsed.  Items (3) and (4) identify 
the eclipsed object as a rotating disk.  Item (5) shows that the
blue -- ultraviolet flux level was depressed during the eclipse so
that nebular lines became visible.

Apparent confirming proof of eclipses in CH Cyg comes from the prediction 
and then observation of the eclipse of 1999 \citep{eyres_et_al_2002,
sokoloski_kenyon_2003b}.  \citet{eyres_et_al_2002} presented $UBV$
photometry for the 1999 eclipse, revealing that the eclipse is not visible 
at $V$, but is clearly detectable in $B$ and much more pronounced in
$U$.  This is the expected enhancement of an eclipse of a hot object by a
cool star.  In addition, they find that flickering is again absent
during the 1999 eclipse, just as in 1985.  The strengths of various
accretion related spectral lines decreased during the eclipse.  
From Figure~5 of \citet{skopal_et_al_2007} the mid-date of the eclipse
in 1999 is estimated as JD 2,451,425.

A major complication to this recent eclipse identification 
is that the 1999 eclipse date, JD 2,451,425, is 550 days earlier than 
predicted by the ephemeris of \citet{mikolajewski_et_al_1990} and 617 days
earlier than our ephemeris.  These differences correspond to an orbital
phase shift of $\sim$0.1.  In eclipses of symbiotic stars
the observed eclipse is not that of the white dwarf, which is
unobservable at visual wavelengths, but rather of a UV bright, hot
spot.  Three dimensional simulations of the
symbiotic-recurrent nova RS Oph by \citet{walder_et_al_2008} predict
an accretion disk that has a diameter 0.1 times the major axis of
the binary system.  This accretion disk is surrounded by a much
larger Archimedean spiral, so there is a large area over which the
hot spot can occur.  From observations of several symbiotics 
\citet{skopal_1998} concluded 
that during quiescent phases of the systems, the times of eclipses
occur prior to the time of inferior conjunction of the giant.  When
the systems are active eclipses coincide with conjunctions predicted
by spectroscopic orbits.  For CH Cyg the \citet{mikolajewski_et_al_1990}
ephemeris from active periods agrees with the times of inferior
conjunction given in Equation 1.  The 1999 eclipse took place during
the period of declining activity preceding the current extended
period of quiescence \citep{eyres_et_al_2002,skopal_et_al_2007}.
We suggest that the accretion disk was undergoing changes at the
time of the 1999 eclipse.

The details of the light curve shape and hydrogen line profiles in
these eclipses add further support to the eclipse interpretation.
In all cases the $U$ light is brighter before the eclipse than after.
In cataclysmic systems this is a well-known eclipse hallmark.
The pre-eclipse brightening results from the accreting side of
the disk facing the mass donating star and rotating in the direction
of orbital motion.  For H$\alpha$ and H$\beta$ line profiles, observed
through eclipse \citep{eyres_et_al_2002, fernie_et_al_1986},
the negative side of the profile is eclipsed first and reappears
first.  Thus the accretion disk is rotating with the side moving
away from the red giant on the leading side of the orbit.

The eclipses discussed above are associated with the 15.6 yr 
orbit.  Yet another complication in the discussion of CH~Cyg eclipses
is that \citet{skopal_1995}, \citet{skopal_et_al_1996}, and subsequent
papers by Skopal claim that there are also eclipses in the 2.1 yr 
orbit.  \citet{skopal_1995} reported the presence of $U$-band decreases 
spaced at $\sim$2.1 year intervals on the light curve.  Supporting evidence
is seen in the disappearance of flickering during at least some of
these events.  Counter to the 2.1 yr eclipse claim  
\citet{mikolajewski_et_al_1990b} and
\citet{sokoloski_kenyon_2003b} found a general correlation of the
strength of the flickering with the luminosity of the blue-violet
continuum.  \citet{sokoloski_kenyon_2003a} suggested that the events
observed by \citet{skopal_et_al_1996} are not eclipses but result
from collapses of the inner accretion disk.

\subsubsection{Flickering and Jets}

Flickering is an indicator of a compact mass-accreting object in
the system.  \citet{mikolajewski_et_al_1990b} found that the rapid
optical variations (flickering) of CH Cyg have a period of $\sim$500~s.
There is a range in the periods reported for optical flickering in
CH Cyg, e.g., \citet{hoard_1993} and \citet{rodgers_et_al_1997}
suggested a period in the range $\sim$ 2200 -- 3000 s \citep[for a
detailed discussion see][]{sokoloski_kenyon_2003b}.  With any of
these time scales the flickering must originate at a hot component
in the CH Cyg system since the dynamical time scale for the giant,
t$_{dyn}$$\sim$$(R^3/2GM)^{1/2}$, is on the order of a month.
Flickering is also seen in X-ray observations, confirming the white
dwarf nature of the secondary \citep{ezuka_et_al_1998}.

CH Cyg is known to have both radio \citep{taylor_et_al_1986} and
optical \citep{solf_1987} jets, which are approximately in the
plane of the sky.  This provides a first limit on the inclination
of the orbit, although the information from the eclipses is far
more accurate.  \citet{crocker_et_al_2002} found from a time series
of observations that the jet is precessing with a period $\sim$18
$\pm$ 0.5 yr.  While the precession mechanism is uncertain, the
period of the precession is similar to the 15.6 yr orbit, suggesting
that the white dwarf is in that orbit.  

\citet{sokoloski_kenyon_2003a} discussed the relation between the
inner disk and the development of jets.  They argued that it is
impossible to understand the jet and optical activity of CH Cyg
without an accretion disk around a white dwarf companion.
\citet{sokoloski_kenyon_2003a} pointed out a strong correlation between
optical flickering in CH Cyg and changes in flux.  The decrease or
disappearance of flickering corresponds to a decrease in the blue
flux.  They found that this is related to disruption of the inner
disk and follows episodes of jet production.

\subsection{Distance, Luminosity, and Radius}

\subsubsection{Giant}

\citet{kenyon_fernandez_1987} undertook a detailed photometric
calibration of symbiotic spectral types.  From the resulting 
type of M6.5 ($\pm$ 0.3) III for CH Cyg, \citet{murset_et_al_1991}
used a spectroscopic parallax to derive a distance of 240 ${+30 \atop
-20}$ pc.  This is in excellent agreement with the 268$\pm$66 pc
distance based on Hipparcos observations \citep{viotti_et_al_1997},
which has been used extensively in the literature since
1997.  Recently, \citet{van_leeuwen_2007} reanalyzed the Hipparcos
parallaxes resulting in a distance for CH Cyg that is nearly 10
percent closer, 244 ${+49 \atop -35}$ pc, than the
\citet{viotti_et_al_1997} distance.  The uncertainties for the original 
and revised Hipparcos distances have large overlap.

\citet{biller_et_al_2006} provided a fit to optical and infrared
photometry of CH Cyg, using a Kurucz model spectrum for an M6 III
star and a dust model.  They found a luminosity of 6900 L$_\odot$.
This is in good agreement with the value given by \citet{skopal_1997}
from an energy distribution encompassing the UV through the far-IR.  He
determined a bolometric luminosity of $\sim$8000 $\pm$ 4000 L$_\odot$
for a distance of 268 $\pm$ 66 pc.  This luminosity corresponds to a
radius of $\sim$310 R$_\odot$.

\citet{dyck_et_al_1998} measured a 10.4 mas diameter for CH Cyg at
2.2 $\mu$m.  Assuming the 244 pc distance, the stellar radius
is then 273 R$_\odot$.  \citet{schild_et_al_1999} find a
radius of 280 $\pm$ 65 R$_\odot$, based on near-IR photometry.  The
\citet{dyck_et_al_1998} radius can also be calibrated to an effective
temperature of 3084 $\pm$ 130 K.  This compares 
very well with the \citet{richichi_et_al_1999} 
calibration for spectral type M 7 of 3150 $\pm$ 95 K.

CH Cyg has a variable $K$ mag ranging from $\sim$$-$0.9 to $-$0.3 
\citep{munari_et_al_1996}.  We adopt the
mean as a typical magnitude, $K$ = $-$0.6 and $J-K$ = 1.6.  
\citet{taranova_shenavrin_2007} reported $K$ band fading, with $K$ as
faint as +0.35 and $J-K$ = 2.06, but this appears to be related to dust
formation episodes.  With the 244 pc distance, $K$ = $-$0.6 
corresponds to an absolute $K$
magnitude of $-$7.5 $\pm$ 0.4.  The $K$ band bolometric correction can be
computed from the $J-K$ color, following \citet{bessel_wood_84}, to yield
a bolometric magnitude of $-$4.3 and a luminosity of 4200 $+2000
\atop -1500$  L$_\odot$.  Assuming a temperature of 3100 K, the
effective temperature -- radius -- luminosity relation gives a radius
of $\sim$224 R$_\odot$.

The values of the M giant's basic parameters, derived from the various 
analyses discussed above, are in reasonable agreement.
For the following discussion we adopt values of L $\sim$5000
L$_\odot$, T $\sim$3100 K, and R $\sim$280 R$_\odot$.  The
corresponding bolometric magnitude is $-$4.5.  We note that typical 
radii for other giant stars in symbiotic binaries of earlier spectral
types are 100 - 200 R$_\odot$ \citep{fekel_et_al_03}.

\subsubsection{White Dwarf}

While the white dwarf in the CH Cyg system has not been observed
directly, some properties can be inferred from the accretion process.
As summarized by \citet{karovska_et_al_2007}, X-rays from CH~Cyg
have been detected by a number of groups.  The X-rays originate
from the accretion disk -- white dwarf boundary layer as well as
from a shock, where the jet interacts with circumbinary material.
\citet{ezuka_et_al_1998} uses X-ray spectroscopy to determine the
temperature and bolometric luminosity 
of the white dwarf.  The luminosity derived,  $\sim$10$^{33}$ ergs
s$^{-1}$ ($\sim$ 0.25 L$_\odot$),
can be combined with a mass-radius relation to
set a lower limit on the mass of the white dwarf of 0.44 M$_\odot$.

\subsection{Rotation and Line Widths}

\citet{schmutz_et_al_1994}, \citet{murset_et_al_2000}, and \citet{zamanov_et_al_2007}
argue that
in most S-type symbiotics the giant star is synchronously
rotating.  In the case of synchronous rotation the projected
rotational velocity of the late-type giant can be used to estimate
the stellar radius \citep[e.g.,][]{fekel_et_al_2008}.  The $v~sin~i$ value
for the M giant can be estimated from the full-width at half-maximum
(FWHM) of unblended absorption lines in the near-infrared spectra.
Using atomic lines near 2.223 $\mu$m, we apply the analysis technique
of \citet{fekel_1997} to CH Cyg.  For CH Cyg
an empirical calibration with intrinsic line broadening of 3
km s$^{-1}$ yields $v~sin~i$ of 8 $\pm$ 1 km s$^{-1}$.

Both the 15.6 yr and 2.1 yr orbits of CH Cyg are eccentric.  In
an eccentric orbit the rotational angular velocity of the M giant
synchronizes with the orbital angular velocity at periastron.  This
was called ``pseudo-synchronization'' by \citet{hut_1981}.  Using
equation (42) of \citet{hut_1981}, we calculate a long pseudo-synchronous
period of 5219 days and a short pseudo-synchronous period of 453
days.

Assuming $sin~i$ = 1, the stellar equatorial rotational velocity is
8 km s$^{-1}$.  Combining this value with the pseudo-synchronous
periods \citep{fekel_et_al_03} suggests a stellar radius of 825
R$_\odot$ for the 15.6 yr orbit or 72 R$_\odot$ for the 2.1 yr 
orbit.  The two computed radii are in poor agreement with the 
$\sim$280 R$_\odot$ value from the direct measurement of the stellar
angular diameter.  Thus in both the long and short period cases the
rotation of CH Cyg is not pseudo-synchronous.  
Other symbiotic systems with orbital periods near that of the short 
period are synchronized \citep{fekel_et_al_03, zamanov_et_al_2007}.

In Paper I we found that for CH Cyg lines of modest strength the FWHM
varies by $\sim$ 30\% over the 2.1 yr period.  The pseudo-synchronous
period differs by nearly a factor of two from the spectroscopic
period.  However, the line width modulation is clearly occurring with the
2.1 yr period.  The rotational broadening would
be more than twice that observed for a period of 2.1 yr and a
diameter equal to that of CH Cyg.  We conclude
that the variation in the line widths is not due to a rotational
modulation associated with star spots or plages.

\subsection{Masses}

The values for the stellar parameters of CH~Cyg can be used with 
stellar evolution theory to place additional constraints on mass.  
\citet{schmidt_et_al_2006} determine a solar iron abundance for 
CH~Cyg.  From Table (2)  of \citet{vassiliadis_wood_93} the solar 
abundance and a bolometric magnitude of $-$4.5 for CH Cyg yield 
an initial mass for the M giant between $\sim$1.5 and 3.5 M$_\odot$.  
Other arguments also suggest a mass near 2 M$_\odot$.  If CH~Cyg 
is currently undergoing a helium-shell flash (\S4.1), the luminosity 
is currently near a peak, which requires a mass near the bottom of 
the mass range.  Based on the absolute magnitude and effective 
temperature, \citet{schmidt_et_al_2006} find log~$g$ $\sim$0 for 
the surface gravity of CH Cyg.  This also is in agreement with a 
mass of 2 M$_\odot$ for a radius of 280 R$_\odot$.  

If we assume that the white dwarf  
is not the result of a pathological evolutionary
process, then the relation of \citet{kalirai_et_al_2007} can be
used to map the white dwarf initial to final mass.  The initial mass of the
white dwarf, which evolved from the more massive star in the binary, had to 
be larger than the initial mass
of the current M giant requiring the mass of the white
dwarf to be $>$0.56 M$_\odot$. 

\subsection{Mass Loss}

Gas and dust mass loss rates for a number of symbiotics, based on
IRAS photometry, are given by \citet{kenyon_et_al_1988}.   
Revising the distance to 244 pc, the dust mass loss rate of CH Cyg
is 4 $\times$ 10$^{-7}$ M$_\odot$ yr$^{-1}$ and the gas mass loss
rate is 1 $\times$ 10$^{-8}$ M$_\odot$ yr$^{-1}$.  A mass loss rate
for dust, based on near-IR photometry, is computed by
\citet{taranova_shenavrin_2004} and \citet{taranova_shenavrin_2007}.
They obtain $\sim$3 $\times$ 10$^{-7}$ M$_\odot$ yr$^{-1}$ in
good agreement with \citet{kenyon_et_al_1988}.  A dust formation
episode in late 2006 resulted in multiple magnitude dimming
in the $J$, $H$, and $K$ bands plus dimming of a few tenths
of a magnitude at $L$ and $M$.  Assuming a spherical symmetry for
the dust shell, \citet{taranova_shenavrin_2007} translated the brightness
change into 
a mass-loss rate of 2 $\times$ 10$^{-5}$ M$_\odot$ yr$^{-1}$.

\citet{kenyon_et_al_1988} derived a dust temperature
of 400 K, considerably less than the value of 750-800 K
given by \citet{taranova_shenavrin_2004}.  The radius of the dust
scales inversely to the dust temperature \citep{kenyon_et_al_1988}.
From \citet{kenyon_et_al_1988} equation (5), R$_d$ $\sim$ 108 AU
for 400 K dust and 19 AU for 800 K dust.  The semi-major axis of
the 15.6 yr orbit is on the order of 8 AU, so the dust at either
temperature lies well outside the stellar orbits.

Mid-infrared imaging shows a Gaussian dust shell of $\sim$45 AU
FWHM \citep{biller_et_al_2006}.  This confirms the circumbinary
nature of the dust and suggests that the dust temperature is between
the \citet{kenyon_et_al_1988} and \citet{taranova_shenavrin_2004}
values and in the same temperature range as found in D-type systems
\citep{kotnik_et_al_2007}.  Because the dust is circumbinary,
counter to the assumption of \citet{taranova_shenavrin_2007},
extinction events are no doubt due to localized condensations
(clouds) rather than large scale mass ejection events.

\subsection{Inclination}

\citet{fekel_et_al_2008} note that the minimum inclination for
symbiotic eclipses is given by 
\begin{equation}
cos(i) < (R_{rg} + R_{wd})/a,
\end{equation} where $a$ is the
semimajor axis of the binary, $i$ is the orbital inclination, and $R$
is the radius of the red giant and white dwarf components.  From
\S4.2.1, R = 280 R$_\odot$.  
From the orbital elements the red giant
semimajor axis can be determined, but it is the total semimajor axis 
that is required for equation (2), so a total mass must be assumed.  
For a first estimate we adopt the total mass from stellar evolution 
(\S4.4) of 2.5 M$_\odot$.  Then from Kepler's third law 
and the 15.6 yr period
the semimajor axis
is 8.5 AU resulting in a minimum inclination of 81$^\circ$.  With 
this estimate for the inclination, $sin~i$, which is critical to 
converting the observed mass function into component masses, is 
equal to 1.0 to better than 2\%.

The length of the eclipse gives more precise information about the
inclination.  The eclipse duration is $\sim$ 200 days
\citep{mikolajewski_et_al_1987}. Ingress/egress is short compared
to the length of the eclipse as expected if the size of the eclipsed
UV bright spot is much smaller than the diameter of the red giant.
The 15.6 yr orbit has a small eccentricity of 0.122, and so for
simplicity we assume the orbit is circular.  The eclipse duration is 
then 3.5\% of the orbital period.  For a semimajor axis of 8.5 AU the 
white dwarf moves 404 R$_\odot$ during an eclipse.  This value sets 
a lower limit on the diameter of the red giant.

We can compute the inclination by comparing the distance moved during
the eclipse with the 280 R$_\odot$ radius of the giant.   
From geometry, the white dwarf traverses
a path 194 R$_\odot$ off the center of the M giant, giving an inclination
of 84.0$^\circ$.  This is in agreement with an estimate by
\citet{skopal_1995} of i $>$ 83$^\circ$, based on the shape of the
1985 eclipse and the assumption of R$_{rg}$ = 200 R$_\odot$.

\subsection{Summary of Basic Parameters}

From the above sections we can distill a profile of the components
of CH Cyg.  The M giant velocities show the existence of both a  
short period (2.1 yr) and long period (15.6 yr) ``orbit''.
However, only two stars have been identified directly or indirectly in
the CH Cyg system, an M giant and a white dwarf.  The M giant has $L$
$\sim$5000 L$_\odot$ (M$_{bol}$ =$-$4.5), $T$ $\sim$3100 K, $R$ $\sim$280 
R$_\odot$, and $M$ $\sim$2 M$_\odot$.  The white dwarf has $M$ $\ga$
0.6 M$_\odot$.  The M giant has solar abundances, is on the
thermal pulsing AGB, and is a radial overtone pulsator 
with a period of $\sim$100 days.
The CH Cyg long period system has an orbital period of 15.6 years, 
eclipses, and has an inclination of 84$^\circ$.
These data are summarized in Table 3.

\section{MODELS OF THE CH CYG SYSTEM}

Several models have been proposed to explain the observed 
properties of the CH Cyg system.  In the following sections we briefly 
review and comment on these models.

\subsection{Miko{\l}ajewski et al. 1987}

From the rather poor radial velocities available at that time, 
\citet{mikolajewski_et_al_1987} concluded that CH Cyg is a binary
with a period $\sim$5700 days and a large orbital eccentricity of 0.55.  
While we have confirmed the period, our radial velocities result in a 
greatly reduced eccentricity.  \citet{mikolajewski_et_al_1987} 
estimated that the system consists of an M6 giant 
with a mass of $\sim$3 $M_\odot$ and a white dwarf companion with 
a mass $\sim$1 $M_\odot$.  The white dwarf is surrounded by an
accretion disk formed from the red giant wind.  Eclipses were
identified in $U$ band light curves, and as a result, the inclination
was set to 90$^\circ$.  

\subsection{Hinkle et al. 1993}

\citet[][Paper I]{hinkle_et_al_93} demonstrated that the CH Cyg
system has two independent sets of velocity variations, a 2.1 year
short period and an $\sim$ 15 year long period.  They note that other 
well known symbiotic systems have orbits of about 2 years and that 
CH Cyg can exhibit large amounts of activity associated with mass 
transfer to the secondary.  This was presumed to imply a small orbital 
separation.  In addition, at this time (early 1990s) the evidence 
for eclipses was weak.  To devise a model similar to other active 
symbiotic systems, it was proposed that the components of the 
symbiotic system reside in the short period orbit.  

Paper I then used the mass function 
\citep{russell_et_al_1955}, 

\begin{equation}
f(m) = (m{_{wd}^3}sin^3i)/(m_{rg} + m_{wd})^2,
\end{equation}

\noindent
where $m_{wd}$ is the mass of the white dwarf, $m_{rg}$ is the mass
of the red giant, and $i$ is the orbital inclination, to derive
limits on the masses and inclination.  Constraining the inclination
by the requirement that the observed jet is nearly in the plane of 
the sky required a very low mass white dwarf.  The third star in the 
long period orbit was assumed to be an unseen late-type dwarf.  The 
model presented in Paper I was based on a moderate inclination.  
As discussed in the present paper, this model is no longer tenable
due to improved knowledge of the component masses and orbital inclination.

\subsection{Skopal et al. 1996}

\citet{skopal_et_al_1996} claimed that both long period and short
period eclipses are present in optical photometry and spectroscopy.
They then proposed a revised three body model.  
\citet{skopal_et_al_1996b} felt that similarities between
the 2.1 year period of CH Cyg and other symbiotic systems 
require that the symbiotic system be in the short
period orbit.  To produce the observable eclipses, the G-K dwarf
in Paper I was replaced with a giant in the long period orbit.
In recognition of the single lined nature of the spectrum,
\citet{skopal_1997} argued that the spectrum of the 
giant in the long period orbit is blended with the M7 III primary, and 
one of the two giants is a bright giant.
\citet{taranova_shenavrin_2004} suggested that  
the giants be nearly identical.

This model has several problems.  \citet{mikkola_tanikawa_1998}
found a triple system with a giant in the outer orbit to be unstable.  
There is no evidence in the spectra for the other bright giant. 
If two giants are present in the system, stellar
evolution requires that they have nearly identical masses.  
In this case the inclination required by the mass function is relatively
small and eclipses from the inner orbit will not occur.

\subsection{Mikkola \& Tanikawa 1998}

\citet{mikkola_tanikawa_1998} proposed that there is an inner close
binary of total mass $\sim$4 M$_\odot$ with a white dwarf of
mass $\sim$1$M_\odot$ in an outer orbit.  The inner orbit is at a 
high inclination with respect to the white dwarf orbit.  Activity on 
the white dwarf is driven by a Kozai resonance which causes large 
eccentricity variations in the inner binary.  In the high-eccentricity 
state, gas would be expelled from the red giant causing activity on the 
white dwarf.  This model explains the long periods of inactivity in CH~Cyg.

The masses of the \citet{mikkola_tanikawa_1998} model are not in
agreement with the derived masses for the system, but the basic model
is potentially viable.  We will return to this point in discussing binary
models for the short period variation.

\subsection{Two Star Models}

\citet{munari_et_al_1996} registered several objections to the
triple star model for CH~Cyg.  These included orbital stability, proposed
masses for the components, and modeling the photometric behavior.
They suggested that the short-period variations are pulsational, 
and hence, the system must be a binary.  However, \citet{munari_et_al_1996}
failed to note that a 2.1 yr pulsation exceeds the fundamental
period.  Thus, if the 2.1 yr period results from pulsation, it can 
be neither a normal pulsation mode of the star nor a beat between 
pulsation modes.

\citet{ezuka_et_al_1998} derived a minimum mass of 0.4 M$_\odot$ 
for the white dwarf, which exceeds the limit imposed in Paper I.  They 
concluded that the value of the mass function determined in 
Paper I is implausible.  While their minimum white 
dwarf mass is indeed inconsistent with the result for the 2.1 yr orbit, 
\citet{ezuka_et_al_1998} failed to note that Paper I
presented both short- and long-period mass functions, and the latter 
value is consistent with the 0.4 M$_\odot$ white dwarf mass limit.  

\citet{schmidt_et_al_2006} briefly reviewed the controversy about
whether the CH~Cyg system is a triple or binary system.  They also
discussed whether the white dwarf is in the inner or outer binary.
They noted the discovery that radial velocity variations of 
multiple-mode semiregular M-giants have periods longer than the
fundamental radial mode \citep{hinkle_et_al_2002}. Indeed, CH~Cyg has a 
strong similarity to these systems, which suggests that the
2.1 yr period in CH Cyg could be a non-radial pulsation mode.  If so, the
long period velocity variation discussed in Paper I must correspond to
the orbit containing the symbiotic components, and the CH Cyg system 
is a binary not a triple system.  This is a viable model for the CH Cyg 
system, and we will examine this possibility more extensively below.

\section{THE CH CYG SYMBIOTIC SYSTEM}

Adopting $sin^3i$ = 1 and using the masses of 2.0 and 0.6 $M_\odot$
that were estimated in \S4.4, we compute from equation (3) a value 
for the mass function of 0.032 $M_{\odot}$.  This value is similar
to our 15.6 yr mass function value of 0.051 $M_{\sun}$ but quite
different from our 2.1 yr mass function value of 0.0015
$M_{\odot}$.  Increasing the white dwarf mass to 0.72 $M_\odot$,
which implies a progenitor mass of 3.3 $M_\odot$
\citep{kalirai_et_al_2007}, produces a match to the 15.6 yr mass function 
value of 0.051 $M_\odot$.  
This mass function value is then in agreement with multiple lines of 
evidence, reviewed above, that require the symbiotic
components to be members of the long period system.  For the above masses
Kepler's third law produces a semimajor axis of 8.7 AU.  Assuming that
the 15.6 yr and 2.1 yr orbits are co-planar and that 
the red giant mass is 2.0 M$_\odot$, the 2.1 yr mass function
requires a mass of $\sim$0.2 M$_\odot$ for a postulated secondary star
in that orbit, a point that we will return to in discussing the nature 
of the short period variation.

The masses of the dwarf and giant are, of course, related but not uniquely
constrained by the 15.6 yr mass function.  From the various limits
set on the values for the masses, the red giant mass is in the range
1.5 $\le$ M$_{rg}$ $\le$ 3.0.  From the long period mass function
the corresponding range for the white dwarf is 0.61 $\le$ M$_{wd}$
$\le$ 0.92.  Significantly, all the values for masses of the white
dwarf derived from the long period mass function are in accord with
evolutionary constraints.  The mass range is sufficiently small to
exclude possible models for CH Cyg with either low or high mass
white dwarfs \citep[see for example][]{luna_sokoloski_2007}.

\citet{iben_tutukov_1996} commented in their extensive paper on the
evolution of symbiotic stars that symbiotic systems are a ``nonuniform
family of wide binaries with actively interacting components that
differ in (a) the nature of their accreting components, (b) reasons
for mass exchange, and (c) the physical mechanisms for their observed
variability.''  This captures the difficulty in understanding CH
Cyg.  Although the system is phenomenologically similar to many
S-type symbiotics, in many respects it is a different type of
symbiotic system.  The orbital period of 15.6 yr is more than three
times that of the next longest S-type symbiotic with a well defined
orbit \citep{fekel_et_al_2007}.  In addition, none of the previously
studied S-type symbiotics have a pulsating AGB star in the system.
The D-type symbiotics R Aqr and {\rm o} Cet are fundamental AGB
pulsators (Miras), while CH Cyg is an overtone AGB pulsator.
Consequently CH Cyg has a lower mass loss rate than the fundamental mode
Miras.  However, in the CH Cyg system the lower mass flow to the
secondary is compensated by the smaller semimajor axis.  The much
less studied symbiotic PU Vul is possibly a similar system
\citep{nussbaumer_vogel_1996}.

The Mira symbiotics R Aqr and {\rm o} Cet are similar to CH Cyg in
some respects. For example, R Aqr has a jet, although its 40.9 year
orbital period is nearly three times that of CH~Cyg.  The R Aqr
system semimajor axis is 13-16 AU \citep{gromadzki_mikolajewska_2008}.
{\rm o} Cet has a period more than ten times longer, $\sim$500 yrs,
with 70 AU component separation \citep{prieur_et_al_2002}.  {\rm
o} Cet exhibits flickering on the secondary with a time scale similar
to that seen on CH Cyg \citep{warner_1972}.  Such flickering is not
a common feature of symbiotic systems.  CH Cyg, R Aqr, and {\rm o}
Cet have all been detected at X-ray wavelengths
\citep{karovska_et_al_2007}.  No doubt the comprehensive information about
these three objects, including the X-ray flickering, is also a measure of 
the detail in which these nearby objects can be studied. X-ray flickering
could well be present in other systems but is masked by other X-ray emission 
from the systems or by the larger distances to other systems.

It is interesting to compare the near-IR colors of CH Cyg, both outside
and during dust formation events, with the colors of symbiotic
(D-type) Miras.  The spectral type of CH Cyg is similar to that of many 
Miras near maximum light.  \citet{whitelock_1987} provides infrared
photometry for a selection of symbiotic Miras.  When quiescent, CH
Cyg occupies the domain of normal Miras in a (J-K) -- (K-L) relation.
During a dust formation episode, the CH Cyg colors evolve along the 
line occupied by symbiotic Miras.  Of the symbiotic systems studied by
\citet{whitelock_1987} CH Cyg is most similar to Hen 2-38.

\citet{whitelock_1987} found it unlikely that symbiotic
systems that currently contain Miras would have been recognized as
symbiotic prior to the onset of the Mira high mass-loss phase.  
She also noted, as did
\citet{iben_tutukov_1996}, that there is not an evolutionary relation
between symbiotics.  Symbiotics are binary systems of different separations,
passing through an evolutionary phase that results in mass transfer.

A question frequently raised about CH Cyg is the cause of the changes
in the activity state of the symbiotic system.
As discussed by 
\citet{mikolajewski_et_al_1990} CH
Cyg was quiescent from the first observations in 1885 until 1963.
The interval since 1963 has been dominated by periods of activity 
mixed with quiescent periods.  With a 15.6 year binary orbit 
and normal M giant mass loss, \citet{mikolajewska_et_al_1988} found 
that the Bondi-Hoyle accretion onto a $\sim$1 $M_\odot$
white dwarf from the giant wind was lower by about an order of
magnitude than the accretion required to drive the hot component
luminosity. 

A solution is that widely separated, high mass-loss rate symbiotic
systems are undergoing {\it wind} Roche-lobe overflow \citep[see
for instance][]{podsiadlowski_mohamed_2007}.  The stellar wind in
these systems is directed by the Roche lobe and the efficiency of
mass transfer to the secondary is much higher than might otherwise
be expected.  Models suggest mass transfer efficiencies approaching
100\% instead of a few percent as previously believed.  This addresses
the active states but not the periods of inactivity.  The latter
have been attributed by \citet{sokoloski_kenyon_2003b} to the
collapse of the inner accretion disk.  We will discuss the activity
of CH Cyg below in relation to the nature of the 2.1 yr variation.

\section{THE 2.1 YR PERIOD}

In Paper I we assumed that both the 2.1 yr photometric and spectroscopic
changes were due to orbital motion.  With the recognition that the
symbiotic system resides in the 15.6 yr orbit, we seek an 
understanding of the 2.1 yr periodic variation.

Semiregular variables obey a period-luminosity relation
\citep{wood_sebo_1996}.  From MACHO data a number of overtone series 
were identified in the LMC by \citet{wood_et_al_1999}.  Since the
publication of this seminal work, the relations have been refined
with data from other surveys, most recently those of OGLE
\citep{soszynski_et_al_2007}.  Linear pulsation periods calculated
by \citet{wood_2007} can be compared with the LMC data to identify
the modes of pulsation.  Period-luminosity sequences are present for 
fundamental through fourth overtone pulsation.

CH Cyg is well established as a semiregular variable with a period of
$\sim$100 days.  Its absolute K magnitude is $-$7.5 $\pm$ 0.4 (\S 4.2.1).
To compare CH~Cyg with the LMC relations found by
\citet{wood_et_al_1999} and others, we take a distance modulus of
18.58 from \citet{szewczyk_et_al_2008} for the LMC.  The pulsation mode(s) 
of CH~Cyg then can be identified from the appropriate period-luminosity 
relation.  The CH~Cyg 100 day period and magnitude are within 0.2 
mag of the relation for first overtone pulsation (Fig.~ 4).  This is 
well within the $\sim$0.5 magnitude natural width of the relation
\citep{soszynski_et_al_2007}.

In \S4.1 the 2.1 yr period was identified as a ``long secondary
period'' (LSP).  LSP variables are so named because they all also
have shorter period pulsation.  LSP variables were found by
\citet{wood_et_al_1999} to obey a P-L relation (his sequence ``D'').
The CH Cyg 2.1 yr period is $\sim$0.3 magnitude off the center
of the D sequence (Fig.~4) but again well within the natural width 
of the relation.  \citet{wood_et_al_1999} pointed out that the
range in the ratio of the long to the short period is a function
of the long period and at periods near 1000 days ranges from 5 to
15.  The long to short period ratio for CH Cyg is $\sim$8.
\citet{derekas_et_al_2006} found that no normal stars fall between
the fundamental and LSP sequences.  Figure 4 illustrates that the
2.1 year CH Cyg period is clearly {\it not} fundamental
pulsation.

While approximately 25 -- 30\% of all pulsating AGB stars show LSP
behavior \citep{wood_et_al_1999, percy_et_al_2004}, there is
considerable disagreement about the physical cause.  Possible origins
of the LSPs have been investigated by \citet{hinkle_et_al_2002},
\citet{olivier_wood_2003}, \citet{wood_olivier_kawaler_2004}, \citet{derekas_et_al_2006}, and
\citet{soszynski_2007}, among others.  Six causes have been proposed:
(1) radial pulsation, (2) non-radial pulsation, (3) 
a low mass companion, (4) rotating spheroidal shape for the
star as the result of common-envelope evolution, (5) circumstellar
dust clouds, and (6) star spots.  Of this list, two causes of the LSPs
remain viable after detailed investigations by the above authors:
orbital motion from a close, low-mass companion and non-radial
pulsation.  \citet{wood_olivier_kawaler_2004} concluded that 
the most likely explanation for the LSPs is a low degree $g$ mode
in the outer radiative layers of the AGB star.  However,
\citet{soszynski_et_al_2007} and others noted that the sequence
D variables form an extension of the ellipsoidal variable sequence.
This suggests that at least some LSPs result from extremely
low-mass contact binaries.

CH Cyg appears on the \citet{payne-gaposhkin_1954} and \citet{houk_1963}
lists of LSP variables.  While these authors knew that the long to
short period ratio was in the range 5 to 15, the discovery that AGB
variables have a period-luminosity relation did not occur for
another 25 years \citep{feast_et_al_1989}, and the identification of the
LSP period-luminosity relation followed 10 years later
\citep{wood_et_al_1999}.  Thus, it is not surprising that the 2.1 year
period in CH Cyg has not been extensively investigated as a possible
LSP.  However, it turns out that some of the same arguments that have
been investigated as possible causes of LSP variability
have been investigated for CH~Cyg.
Based on the specifics of the 770 day photometric variability,
\citet{mikolajewski_et_al_1992} concluded that the 770 day period
is not caused by variable extinction, as previously proposed by
\citet{mikolajewski_et_al_1990}.  \citet{mikolajewski_et_al_1992}
also dismissed rotation as a cause since the rotational velocity
required is $\ga$ 13 km s$^{-1}$ and would produce noticeable
line broadening.  In \S4.3 the CH Cyg line width was shown to be
slightly more than half this value.  \citet{mikolajewski_et_al_1992}
hypothesized that the 2.1 yr period might be the result of spots
related to cyclic changes in convective cells.  This origin for LSP
behavior was ruled out by \citet{wood_olivier_kawaler_2004}.

\subsection{Binary Hypothesis}

\citet{soszynski_et_al_2004} and others show that the sequence of
ellipsoidal variables (sequence E) overlaps Wood's sequence D.  The
period-luminosity relation requires that binary systems causing D
sequence photometric variations be contact binaries
\citep{derekas_et_al_2006}.  Based on the shape of the light and
velocity curves, it is apparent that the short period system in
CH~Cyg is not a standard contact binary.  Velocity curves for various
contact binary systems can be found in \citet{lu_rucinski_1999} and
\citet{wood_2007}.  Contact binary radial velocity curves have a
characteristically sinusoidal shape because their orbits are circular.
\citet{adams_et_al_2006} found the sequence E objects in their
sample to be first ascent, giant branch objects with very small
envelope masses, probably as the result of a common envelope mass
ejection event.  However, \citet{soszynski_2007} noted that if the
Roche lobes are not full, it is possible to transfer mass driven
by a stellar wind.  In such a situation the eccentricity can be
increased.  Thus, it is of interest to discuss the binary option
for CH Cyg.

In our discussion of the symbiotic CH Cyg binary system, a number
of arguments have been used to assign mass, effective temperature,
and radius to the M giant (\S4).  In addition, the inclination of
the 15.6 yr orbit is known from eclipses to an uncertainty of less
than a degree.  We initially assume the simple case of coplanar
15.6 yr and 2.1 yr orbits.  It is then possible to solve
for the mass of a possible 2.1 yr companion.  Assuming a 2 $M_\odot$
primary, a putative 2.1 yr companion would have mass 0.2 $M_\odot$.
Using the mass - spectral type calibration of \citet{baraffe_chabrier_1996}
the hypothesized low-mass companion is then an M3 dwarf with an
effective temperature $\sim$3200 K.  This is so similar to the much
more luminous M giant that such a dwarf would be undetectable
spectroscopically.  The Roche lobe radii of a 2 M$_\odot$
and 0.2 M$_\odot$ binary are 260 $R_\odot$ and 96 $R_\odot$
respectively in a circular orbit.  However, the eccentricity $e$
of the 2.1 yr orbit is 0.33.  Thus, the Roche lobe will be constantly
changing from (1+e)a for apastron to (1-e)a for periastron.  The
radius of the giant is $\sim$280 R$_\odot$, so such a system is
Roche lobe filling, and so, at each periastron passage extreme mass
loss would be expected.

In Paper I, instead of a main sequence red dwarf, a {\it white dwarf} 
companion of approximately 0.2 $M_\odot$ was hypothesized.  We now 
argue that a 0.2 $M_\odot$ {\it white dwarf} 
companion can not exist.  The presence of 2 $M_\odot$ star in
the CH~Cyg system requires that any progenitor of a white dwarf
was initially more massive than 2 $M_\odot$.  Such stars
produce white dwarfs that are more massive than 0.6 $M_\odot$
\citep{kalirai_et_al_2007}.  However, we note that 0.2 $M_\odot$ white
dwarfs do exist; \citet{liebert_et_al_2004} reported
the discovery of such an object.  Low mass white dwarfs 
are believed to result from mass transfer and common envelope
evolution with a degenerate companion.  There seems no need
to invoke this 
evolutionary path in the CH~Cyg system.
Even if such a white dwarf companion could exist, a contact
system with a AGB star would be exceedingly active.  In addition,
the X-ray observations of CH~Cyg result in mass limits which require
a more massive white dwarf (\S4.2.2).

\citet{soszynski_et_al_2007} finds that
the binary star explanation of the LSP requires the radius to
semimajor axis ratio to be $\sim$0.4.  Kepler's third law then 
requires a mass of 6 M$_\odot$ for the dwarf.  However,  masses
larger than 2 M$_\odot$ are excluded if the stars in the system
have equal ages.  If the mass of the dwarf and the giant are
equal the mass function requires an inclination of $\sim$8$^\circ$.
This in turn excludes the 
inclined orbit Kozai resonance model of \citet{mikkola_tanikawa_1998}.  
On the other hand, a coplanar result is also evidence that the 
triple model is unphysical because triple systems are seldom if ever
coplanar \citep{muterspaugh_et_al_2008}.

The 2.1 yr velocity variation (Fig. 3) is not sinusoidal.  This 
provides one of the most powerful arguments against the 2.1 yr binary 
hypothesis.  In fact, the eccentricity of 0.33 is moderately large.  
Polar views of the 15.6 yr and 2.1 yr orbits are shown in Fig.~5.
\citet{soker_2000} provides formulae for computing the time scale
for tidal synchronization.  Because the low mass companion does not
have sufficient mass to spin up the envelope of the M giant, the
synchronization time scale is the merger time scale,
$\sim$1000 years \citep{wood_olivier_kawaler_2004}.  The
time scale for orbital circularization is longer than the time scale
for synchronized rotation.  Merger in the synchronization time scale
makes this argument mainly academic.  However, the circularization
time scale is also short, $\sim$5000 years, so all contact binary
orbits of this type would be expected to be circular.
As noted by \citet{wood_olivier_kawaler_2004}, there are
many known LSP systems, so eccentric orbits, even if possible 
for one object, could not possibly be the case for the entire class.

\citet{soszynski_2007} suggested that mass lost by the red giant in a
binary system with a low mass star follows a spiral pattern
producing the LSP variations.  Such spiral mass loss could then
affect the symbiotic binary system.  However, as discussed in \S4.5
infrared observations of CH Cyg require a dust temperature which
places the dust well outside the the 15.6 yr long period orbit.
Furthermore, \citet{mikolajewski_et_al_1992} find that the 770 day
variation in CH Cyg, while color dependent, is inconsistent with
reddening.

These arguments provide convincing evidence that the 2.1 yr period 
variation of CH Cyg does {\it not} result from binary motion.  For the 
entire LSP group \citet{derekas_et_al_2006}
used arguments based on a multivariate test of the amplitude-luminosity
relation to conclude that the LSP phenomenon is in general not a result of
contact binaries.

\subsection{Pulsation}

While it has been shown theoretically that radial pulsation can not
be a cause of the LSPs, this is still frequently listed among the
possible causes.  In the case of CH Cyg several authors have suggested
that the 770 day period is the result of radial pulsation of the M
giant \citep[see][]{munari_et_al_1996}.  We emphasize that a period
of 770 days can not possibly be a radial pulsation.  The frequencies
of radial pulsations in AGB stars can be calculated as a function
of bolometric magnitude \citep{hughes_wood_1990, wood_2007}.  The
fundamental period as a function of absolute $K$ magnitude is plotted
on Fig.~4.  At the luminosity of CH~Cyg, $K$ = $-$7.5 mag, the radial 
fundamental period is $\sim$250 days.  If the origin of the LSP is 
pulsational, the pulsations must be non-radial in nature.

Spectroscopic studies of LSPs have been carried out by
\citet{hinkle_et_al_2002}, \citet{olivier_wood_2003}, and
\citet{wood_olivier_kawaler_2004}.  There are two remarkable qualities
of the LSP velocities.  First, the radial velocity variations among
LSP stars are nearly identical.  The velocity phase curves for seven
LSP stars from \citet{hinkle_et_al_2002} and
\citet{wood_olivier_kawaler_2004} are shown in Fig.~6.  The position
of these objects on the period-luminosity relation is shown in
Fig.~4.  Although the original orbit for one of these stars, AF~Cyg,
had $e$ = 0.08 $\pm$ 0.20, given the large uncertainty, we have
recomputed the orbit with $e$ = 0.3 as shown in Fig.~6 and found a
very acceptable fit to the velocities, although the computed velocity
curve is in antiphase with those of the other stars.  The 2.1 yr
velocity phase curve of CH Cyg matches those of the other LSP stars.
The remarkable similarity of the ``orbits'' is inconsistent with
both the expectation of random orbital orientation as well as the
circularization constraints discussed above.  The variables included
in Fig.~6 are large visual amplitude LSP systems, so the similarity
of velocity amplitudes could be a selection effect.  

Second, the LSP velocity variations totally dominate the shorter
pulsation period in the LSP objects studied.  For instance, in CH
Cyg the overtone pulsations are not detectable in the velocities
(\S3) and result in only a low amplitude photometric variation (\S4.1).
Thus, LSP velocity changes are not easily differentiated from orbital
motion for stars where the photometric variation is poorly known.
This could be pervasive among suspected binary systems.  A possible
example is the velocity curve of the X-ray, suspected neutron star
symbiotic binary HD 154791 \citep{galloway_et_al_2002}, which matches
the parameters of the LSP ``orbits'' of the stars in Fig. 6.

As noted by \citet{wood_olivier_kawaler_2004}, velocity curves with
the shape shown in Fig.~6 can be reproduced by rotating prolate
spheroids.  The symmetry of such a spheroid requires a rotation period
twice the period of the velocity curve.  This period is in agreement
with 8 km s$^{-1}$ measured line widths in CH Cyg.   \citet{kiss_et_al_2000}
proposed a rotating pulsating ellipsoid model for the unusual light
curves of two semiregular variables.  A rotating prolate spheroid could
result from a recent binary merger, but it is difficult to understand
how an entire class of variables following a period-luminosity law could
be the result of recent mergers.

The observed properties of a rotating prolate spheroid can be generated 
by low order non-radial pulsation.  In the case of non-radial pulsation a
period-luminosity relation would exist.  Non-radial pulsation can
be prograde, i.e. in the direction of stellar rotation, or retrograde.
\citet{unno_et_al_1989} stated that waves traveling in retrograde,
as opposed to prograde, result in identical velocity curves but with
the sign reversed.  This could explain the antiphase velocity curve 
of AF~Cyg in Fig.~6. 

\citet{hatzes_1996} provides models for some low l sectoral (l = $-$m)
modes.  Unfortunately, the models are only for m = 2, 4, and 6.  For these
modes there are at minimum 4 sectors around the star and the velocity 
variations are nearly sinusoidal.  The velocity amplitude decreases with 
increasing m because for higher m there are a larger number of 
velocity zones with cancelling sign \citep{hatzes_1996}.  The amplitude 
and asymmetry of CH~Cyg suggest an l = 1 mode.  From the relations given by
\citet{hatzes_1996} the l = 1 mode pulsation velocity amplitude is
about half of the observed velocity amplitude.  An interesting
aspect of non-radial modes is that the vertical scale height is
small.  

Additional insight into the LSP problem comes from K to early M
{\it red} giants.  For these stars \citet{henry_et_al_2000} reported
the presence of short period, i.e.\ time scales of a few days to weeks,
radial and non-radial pulsations.  \citet{henry_et_al_2000} found
non-radial pulsations restricted to the hot side of the coronal
dividing line at about K2.  For an early K giant the radial periods
are typically days.  The K giants have secondary periods with
lengths of hundreds of days.  There is considerable uncertainty
about the origin of the LSPs for these stars, as for the M giant
LSPs.  As is the case for AGB LSPs, rotation with star spots, non-radial
pulsations, and low mass companions have been discussed as explanations.
In the case of the K0 III star $\pi$ Her, \citet{hatzes_cochran_1999}
found the pressure scale height for the photosphere to be a factor of 10
smaller than the radial wavelength of the 613 day pulsations.

The hypothesis of non-radial modes in late-type giants has 
been impeded by the knowledge that 
gravity modes are evanescent in convective regions.  This makes
their postulated existence in both AGB and red giants difficult to
understand.  Thus, \citet{hatzes_cochran_1999} suggested r-mode
oscillations \citep{wolff_1996}, which arise in rotating, convective
fluids, as an alternative.  However, this alternative raises a
different problem for AGB stars.  In AGB stars the rotational period
results in a pulsation period which is too long.  For r-modes the
pulsation frequency is equal to the rotation frequency for l = 1 and
longer for higher modes \citep{wolff_1996}.  In the case of CH~Cyg
the rotation period is more than twice the 2.1 year pulsation period.
A solution proposed by \citet{wood_olivier_kawaler_2004} relies on
the point that LSP non-radial modes occur only in stars that also have
radial pulsation.  The proposal is that radial pulsation thickens
the radiative layer above the convective layer allowing the propagation
of g-modes.

\citet{wood_olivier_kawaler_2004} discussed light and color variations
in LSP stars.  These can be compared with the variations in CH Cyg
(\S 4.1) and predictions from stellar pulsation.  The reported
changes in spectral type are consistent with pulsation related
changes in temperature.   The types of variation reported by
\citet{mikolajewski_et_al_1992} are fully consistent with variations
for other LSP stars and with variations expected from stellar
pulsation \citep{wood_olivier_kawaler_2004}.

A complication in understanding the CH Cyg system has been the
report of short period eclipses by \citet{skopal_et_al_1996}.
\citet{wood_olivier_kawaler_2004} concluded that for LSP stars the 
strength of H$\alpha$ is a function of phase.  They interpreted this as
changing chromospheric activity.  At some phases the chromosphere
is absent, while at other phases it covers up to 70\% of the stellar
surface.  This same phenomenon has been reported for the K giant
LSP stars.  For $\alpha$~Boo the strength of He~I 10830 \AA~is
correlated with the LSP \citep{hatzes_cochran_1993}.  We suggest
that the UV changes in CH Cyg correlated with the short period phase
found by \citet{skopal_et_al_1996} and \citet{eyres_et_al_2002} are
pulsation driven changes in the chromosphere.  From the dates of
the deep UV minima given in \citet{skopal_1995} the 2.1 yr 
orbital phase (from the eccentric orbit) is 0.14, which is conjunction
in an orbital model or a time of null velocity movement in a
pulsational model.  The absence of flickering during at least some
of these events could be unrelated changes of structure in the white
dwarf accretion disk \citep{sokoloski_kenyon_2003a}.

In Paper I the change of line shape with short period phase was
discussed.  The spectral lines have the largest FWHM at $\phi$ =
0.4--0.6.  However, the lines are more narrow and symmetric at phase
0.4 becoming more asymmetric at phase 0.8.  For K giants
\citet{hatzes_1996} has noted the change in bisector shape with
phase and mode in non-linear pulsations.  The resolution and
signal-to-noise ratio of the current CH Cyg spectra does not allow
detailed comparison but indicates that further study would be of
interest.

A long standing question for CH Cyg is why this star was an M giant
spectral type standard for the first half of the 20th century and
then became an active symbiotic star.  At the heart of this question
is the the nature of the accretion onto the white dwarf secondary.
A possibly naive expectation is that the LSP variation is a controlling
factor in the M giant mass loss.  We speculate that in the case of a non-radial
pulsation origin for LSP, the non-radial modes are stable but
variations in the dominant mode can occur.  If the dominant mode
controls other stellar properties, for instance the mass loss rate
or the directionality of the mass loss, the mass transfer in a
symbiotic system would  change.  Observational tests employing
samples of LSP variables should be possible.  

\section{CONCLUSIONS}

The near-infrared radial velocities of CH~Cyg conclusively show that 
the M giant in the system exhibits two different velocity variations,
a 15.6 year ``long'' period and a 2.1 year ``short'' period.  We
have reviewed the literature on the basic parameters for the CH~Cyg
giant and have estimated its mass and radius.  Evolutionary arguments
require an M giant mass near 2 $M_\odot$. X-ray observations 
require a white dwarf companion more massive than 0.44 M$_\odot$.
Various lines of evidence but most compellingly observations of 
eclipses demand that the inclination of the 15.6 year orbit is 
nearly edge on.  The long period mass function is then matched by 
the assumption of a 0.7 M$_\odot$ white dwarf.

The CH Cyg symbiotic system is an unusual one.  The separation
between the giant and white dwarf is about four times larger than
that for typical S-type systems.  CH~Cyg is a first overtone
pulsating AGB star, which will drive symbiotic mass transfer at a
larger radius. 
While CH Cyg is classified as an S-type system, there
are a number of similarities with D-type Mira systems.
Changing activity in the CH Cyg symbiotic system 
is possibly related to variations in the short-period activity.  

The observed 2.1 yr velocity variation is indistinguishable from that  
seen in stars with long secondary periods (LSP).  LSP objects
occupy a track on the period-luminosity diagram, identified by
\citet{wood_et_al_1999} as sequence D.  The cause of the LSP
variations, while discussed exhaustively, remains uncertain.  LSP
is the only theoretically unexplained type of stellar variability
\citep{wood_olivier_kawaler_2004}.  In the case of CH Cyg there
seem to be only two viable options to produce the 2.1 yr period. 
The star could be a rotating prolate spheroid, perhaps as the result
of a common envelope phase with a former low-mass companion.  However,
with this model it is hard to understand why there are so many LSP
stars.  Much more likely, CH Cyg is undergoing low order non-radial
pulsations that mimic the observed properties of a rotating prolate
shape.  If these are g-modes, the outer stellar structure of pulsating
M giants does not follow standard models because an outer radiative
layer is required to propagate the g-modes.  While this structure
change is not required for r-modes, the periods of r-modes are
excessively long.

\acknowledgements
This research has been supported in part by NASA grant NCC5-511
and NSF grant HRD-9706268 to Tennessee State University.  We have
made use of the SIMBAD database, operated by CDS in Strasbourg,
France, as well as NASA's Astrophysics Data System Abstract Service.
We thank D. Willmarth and D. Harmer for assisting at the
KPNO coud\'e feed telescope.

\clearpage

\begin{deluxetable}{lcrcccr}
\tablenum{1}
\tablewidth{0pt}
\tablecaption{RADIAL VELOCITIES OF CH CYG}
\tablehead{\colhead{HJD} & \colhead{RV} & \colhead{$O-C$} &
\colhead{} & \colhead{$V_L$} & \colhead{} & \colhead{$V_S$}\\
\colhead{2,400,000 +} & \colhead{(km~s$^{-1}$)} & \colhead{(km~s$^{-1}$)}
& \colhead{$\phi${$_L$}} & \colhead{(km~s$^{-1}$)} & \colhead{$\phi${$_S$}}
& \colhead{(km~s$^{-1}$)}
}
\startdata
 43,913.138  & $-$59.2 &    0.07 &  0.689 & $-$60.49 &  0.493 &     1.37 \\
 43,917.267  & $-$59.2 &    0.13 &  0.690 & $-$60.45 &  0.499 &     1.38 \\
 44,263.275  & $-$65.4 &    0.01 &  0.751 & $-$62.09 &  0.960 &  $-$3.29 \\
 44,297.144  & $-$66.2 & $-$1.63 &  0.757 & $-$63.88 &  0.005 &  $-$3.94 \\
 44,507.607  & $-$60.5 &    0.39 &  0.794 & $-$62.74 &  0.286 &     2.63 \\
 44,622.194  & $-$63.9 & $-$1.99 &  0.814 & $-$65.55 &  0.439 &  $-$0.33 \\
 44,660.373  & $-$62.9 & $-$0.52 &  0.821 & $-$64.22 &  0.490 &     0.80 \\
 44,692.110  & $-$63.6 & $-$0.79 &  0.826 & $-$64.60 &  0.532 &     0.21 \\
 44,974.297  & $-$68.0 & $-$0.03 &  0.876 & $-$64.60 &  0.908 &  $-$3.44 \\
 45,279.621  & $-$63.7 & $-$1.10 &  0.929 & $-$65.89 &  0.315 &     1.09 \\
 45,360.478  & $-$63.6 & $-$0.62 &  0.944 & $-$65.35 &  0.423 &     1.12 \\
 45,392.184  & $-$61.2 &    1.99 &  0.949 & $-$62.69 &  0.465 &     3.48 \\
 45,475.199  & $-$64.9 & $-$1.00 &  0.964 & $-$65.53 &  0.576 &  $-$0.37 \\
 45,490.162  & $-$64.6 & $-$0.55 &  0.966 & $-$65.05 &  0.596 &  $-$0.10 \\
 45,507.001  & $-$64.2 &    0.03 &  0.969 & $-$64.43 &  0.618 &     0.26 \\
 45,604.862  & $-$66.0 & $-$0.52 &  0.987 & $-$64.70 &  0.749 &  $-$1.82 \\
 45,647.498  & $-$65.5 &    0.64 &  0.994 & $-$63.40 &  0.806 &  $-$1.46 \\
 45,683.282  & $-$67.7 & $-$1.00 &  0.000 & $-$64.90 &  0.853 &  $-$3.80 \\
 45,721.218  & $-$67.2 & $-$0.07 &  0.007 & $-$63.83 &  0.904 &  $-$3.45 \\
 45,752.156  & $-$66.7 &    0.36 &  0.012 & $-$63.26 &  0.945 &  $-$3.07 \\
 45,776.080  & $-$66.6 & $-$0.04 &  0.017 & $-$63.56 &  0.977 &  $-$3.08 \\
 45,782.082  & $-$66.4 & $-$0.04 &  0.018 & $-$63.54 &  0.985 &  $-$2.91 \\
 45,799.055  & $-$65.9 & $-$0.24 &  0.021 & $-$63.66 &  0.008 &  $-$2.48 \\
 45,812.076  & $-$64.9 &    0.13 &  0.023 & $-$63.23 &  0.025 &  $-$1.54 \\
 45,837.034  & $-$64.2 & $-$0.43 &  0.027 & $-$63.67 &  0.058 &  $-$0.96 \\
 45,948.828  & $-$60.2 &    0.36 &  0.047 & $-$62.31 &  0.207 &     2.47 \\
 45,986.495  & $-$61.5 & $-$1.28 &  0.054 & $-$63.75 &  0.258 &     0.97 \\
 46,005.414  & $-$60.3 & $-$0.18 &  0.057 & $-$62.55 &  0.283 &     2.06 \\
 46,018.341  & $-$59.9 &    0.17 &  0.059 & $-$62.12 &  0.300 &     2.39 \\
 46,044.386  & $-$60.3 & $-$0.29 &  0.064 & $-$62.44 &  0.335 &     1.85 \\
 46,064.855  & $-$60.5 & $-$0.51 &  0.067 & $-$62.54 &  0.362 &     1.53 \\
 46,107.908  & $-$60.5 & $-$0.48 &  0.075 & $-$62.27 &  0.419 &     1.28 \\
 46,125.872  & $-$60.5 & $-$0.44 &  0.078 & $-$62.13 &  0.443 &     1.18 \\
 46,157.791  & $-$61.0 & $-$0.85 &  0.084 & $-$62.35 &  0.486 &     0.49 \\
 46,186.830  & $-$61.0 & $-$0.73 &  0.089 & $-$62.06 &  0.525 &     0.33 \\
 46,223.774  & $-$60.0 &    0.46 &  0.095 & $-$60.65 &  0.574 &     1.11 \\
 46,310.529  & $-$61.7 & $-$0.56 &  0.111 & $-$61.16 &  0.690 &  $-$1.10 \\
 46,344.562  & $-$61.3 &    0.21 &  0.117 & $-$60.19 &  0.735 &  $-$0.90 \\
 46,426.835  & $-$63.2 & $-$0.60 &  0.131 & $-$60.53 &  0.845 &  $-$3.27 \\
 46,452.877  & $-$63.2 & $-$0.29 &  0.136 & $-$60.07 &  0.879 &  $-$3.42 \\
 46,478.799  & $-$63.5 & $-$0.43 &  0.140 & $-$60.06 &  0.914 &  $-$3.86 \\
 46,492.844  & $-$63.0 &    0.04 &  0.143 & $-$59.52 &  0.933 &  $-$3.44 \\
 46,509.786  & $-$62.1 &    0.73 &  0.146 & $-$58.74 &  0.955 &  $-$2.63 \\
 46,569.772  & $-$60.2 &    0.26 &  0.156 & $-$58.88 &  0.035 &  $-$1.06 \\
 46,696.608  & $-$56.4 &    0.01 &  0.178 & $-$58.50 &  0.204 &     2.10 \\
 46,721.501  & $-$55.5 &    0.66 &  0.183 & $-$57.72 &  0.237 &     2.88 \\
 46,748.537  & $-$54.7 &    1.31 &  0.188 & $-$56.95 &  0.273 &     3.56 \\
 46,818.466  & $-$56.3 & $-$0.38 &  0.200 & $-$58.32 &  0.367 &     1.65 \\
 46,838.810  & $-$55.7 &    0.26 &  0.203 & $-$57.60 &  0.394 &     2.16 \\
 46,869.183  & $-$56.9 & $-$0.85 &  0.209 & $-$58.58 &  0.434 &     0.83 \\
 46,931.016  & $-$56.1 &    0.27 &  0.220 & $-$57.22 &  0.517 &     1.39 \\
 46,949.766  & $-$55.4 &    1.10 &  0.223 & $-$56.32 &  0.542 &     2.02 \\
 47,053.636  & $-$56.3 &    1.19 &  0.241 & $-$55.86 &  0.680 &     0.76 \\
 47,069.749  & $-$58.1 & $-$0.40 &  0.244 & $-$57.41 &  0.702 &  $-$1.09 \\
 47,113.295  & $-$58.7 & $-$0.38 &  0.252 & $-$57.25 &  0.760 &  $-$1.83 \\
 47,159.503  & $-$59.7 & $-$0.62 &  0.260 & $-$57.36 &  0.821 &  $-$2.96 \\
 47,201.195  & $-$59.7 &    0.04 &  0.267 & $-$56.60 &  0.877 &  $-$3.07 \\
 47,229.269  & $-$59.6 &    0.40 &  0.272 & $-$56.16 &  0.914 &  $-$3.04 \\
 47,286.152  & $-$58.6 &    0.57 &  0.282 & $-$55.86 &  0.990 &  $-$2.17 \\
 47,307.126  & $-$58.9 & $-$0.61 &  0.286 & $-$57.00 &  0.018 &  $-$2.51 \\
 47,344.045  & $-$56.9 & $-$0.35 &  0.292 & $-$56.67 &  0.067 &  $-$0.58 \\
 47,488.413  & $-$53.6 &    0.24 &  0.318 & $-$55.85 &  0.260 &     2.49 \\
 47,540.384  & $-$54.5 & $-$0.63 &  0.327 & $-$56.66 &  0.329 &     1.53 \\
 47,783.636  & $-$55.3 &    0.73 &  0.370 & $-$55.16 &  0.653 &     0.60 \\
 47,985.231  & $-$60.0 & $-$0.57 &  0.405 & $-$56.53 &  0.922 &  $-$4.04 \\
 48,352.289  & $-$55.5 & $-$0.88 &  0.470 & $-$57.31 &  0.412 &     0.93 \\
 48,353.143  & $-$55.3 & $-$0.67 &  0.470 & $-$57.10 &  0.413 &     1.13 \\
 48,377.043  & $-$54.9 & $-$0.04 &  0.474 & $-$56.52 &  0.445 &     1.57 \\
 48,440.068  & $-$56.2 & $-$0.63 &  0.485 & $-$57.23 &  0.529 &     0.40 \\
 48,586.221  & $-$58.6 & $-$0.70 &  0.511 & $-$57.64 &  0.723 &  $-$1.67 \\
 48,644.104  & $-$58.8 &    0.31 &  0.521 & $-$56.77 &  0.801 &  $-$1.72 \\
 48,736.250  & $-$61.8 & $-$0.99 &  0.537 & $-$58.33 &  0.923 &  $-$4.46 \\
 48,958.250  & $-$54.6 &    1.26 &  0.576 & $-$56.77 &  0.219 &     3.43 \\
 49,802.024  & $-$59.6 & $-$0.26 &  0.724 & $-$61.71 &  0.344 &     1.85 \\
 49,803.979  & $-$59.4 & $-$0.04 &  0.725 & $-$61.50 &  0.347 &     2.05 \\
 49,874.957  & $-$58.7 &    1.43 &  0.737 & $-$60.34 &  0.442 &     3.07 \\
 49,923.777  & $-$60.6 &    0.19 &  0.746 & $-$61.80 &  0.507 &     1.38 \\
 49,997.740  & $-$61.9 &    0.05 &  0.759 & $-$62.26 &  0.605 &     0.40 \\
 49,999.695  & $-$61.7 &    0.28 &  0.759 & $-$62.03 &  0.608 &     0.61 \\
 50,000.742  & $-$61.5 &    0.50 &  0.759 & $-$61.82 &  0.609 &     0.82 \\
 50,162.046  & $-$64.7 &    0.67 &  0.788 & $-$62.32 &  0.824 &  $-$1.71 \\
 50,163.043  & $-$64.5 &    0.90 &  0.788 & $-$62.10 &  0.826 &  $-$1.50 \\
 50,320.682  & $-$65.1 & $-$0.20 &  0.815 & $-$63.80 &  0.036 &  $-$1.50 \\
 50,385.709  & $-$63.8 & $-$1.14 &  0.827 & $-$64.97 &  0.123 &     0.03 \\
 50,567.995  & $-$62.0 &    0.33 &  0.859 & $-$64.03 &  0.366 &     2.36 \\
 50,627.951  & $-$62.4 &    0.48 &  0.870 & $-$64.01 &  0.445 &     2.09 \\
 50,687.829  & $-$63.4 &    0.15 &  0.880 & $-$64.45 &  0.525 &     1.21 \\
 50,750.724  & $-$65.0 & $-$0.62 &  0.891 & $-$65.32 &  0.609 &  $-$0.30 \\
 50,932.878  & $-$67.5 &    0.08 &  0.923 & $-$64.72 &  0.852 &  $-$2.70 \\
 50,981.853  & $-$66.9 &    1.33 &  0.932 & $-$63.45 &  0.917 &  $-$2.12 \\
 51,051.770  & $-$66.0 &    0.87 &  0.944 & $-$63.85 &  0.011 &  $-$1.28 \\
 51,106.744  & $-$63.8 &    0.58 &  0.954 & $-$64.06 &  0.084 &     0.84 \\
 51,295.020  & $-$62.3 & $-$0.26 &  0.987 & $-$64.44 &  0.335 &     1.88 \\
 51,345.808  & $-$61.9 &    0.25 &  0.996 & $-$63.76 &  0.403 &     2.10 \\
 51,346.857  & $-$61.7 &    0.45 &  0.996 & $-$63.55 &  0.404 &     2.30 \\
 51,415.767  & $-$61.0 &    1.46 &  0.008 & $-$62.28 &  0.496 &     2.73 \\
 51,477.760  & $-$60.9 &    1.96 &  0.019 & $-$61.51 &  0.578 &     2.56 \\
 51,648.009  & $-$65.6 & $-$0.88 &  0.049 & $-$63.50 &  0.805 &  $-$2.98 \\
 51,677.971  & $-$65.7 & $-$0.56 &  0.054 & $-$63.02 &  0.845 &  $-$3.24 \\
 51,736.808  & $-$65.1 &    0.50 &  0.064 & $-$61.63 &  0.924 &  $-$2.97 \\
 51,831.701  & $-$61.4 &    0.98 &  0.081 & $-$60.60 &  0.050 &     0.18 \\
 54,271.980  & $-$54.2 &    0.51 &  0.510 & $-$56.42 &  0.304 &     2.73 \\
 54,272.956  & $-$54.2 &    0.51 &  0.510 & $-$56.41 &  0.305 &     2.73 \\
 54,592.984  & $-$58.5 &    0.42 &  0.566 & $-$57.43 &  0.732 &  $-$0.65 \\
 54,634.970  & $-$59.7 &    0.13 &  0.574 & $-$57.86 &  0.788 &  $-$1.71 \\
 54,635.982  & $-$59.3 &    0.55 &  0.574 & $-$57.44 &  0.789 &  $-$1.31 \\
\enddata
\end{deluxetable}

\clearpage

\begin{deluxetable}{lcc}
\tablenum{2}
\tablewidth{0pt}
\tablecaption{ORBITAL ELEMENTS AND RELATED PARAMETERS OF CH CYG}
\tablehead{\colhead{} & \colhead{Short-Period} & \colhead{Long-Period} \\
\colhead{Parameter} & \colhead{``Orbit''} & \colhead{Orbit}
}
\startdata
$P$ (days)             & 750.1 $\pm$ 1.3         & 5689.2 $\pm$ 47.0 \\
$P$ (years)            & 2.0537 $\pm$ 0.0036      & 15.58 $\pm$ 0.13 \\
$T$ (HJD)              & 2,447,293.5 $\pm$ 12.9  & 2,445,681 $\pm$ 192 \\
$\gamma$ (km~s$^{-1}$) &  ...                    & $-$59.91 $\pm$ 0.09 \\
$K$ (km~s$^{-1}$)      & 2.87  $\pm$ 0.13        & 4.45 $\pm$ 0.12 \\
$e$                    & 0.330 $\pm$ 0.041       & 0.122 $\pm$ 0.024 \\
$\omega$ (deg)         & 229.5  $\pm$ 7.7        & 216.9 $\pm$ 12.7 \\
$a$~sin~$i$ (10$^6$ km) & 27.90 $\pm$ 12.30      & 345.69 $\pm$ 9.09  \\
$f(m)$ ($M_{\sun}$)     & 0.0015 $\pm$ 0.0002    & 0.051 $\pm$  0.004 \\
\enddata
\end{deluxetable}

\clearpage

\begin{deluxetable}{ccc}
\tablenum{3}
\tablewidth{0pt}
\tablecaption{DERIVED CH CYG COMPONENT PARAMETERS}
\tablehead{\colhead{Parameter} & \colhead{Value} & \colhead{Reference} 
}
\startdata
 & Red Giant: &   \\
$M_{bol}$      & $-$4.5         &  \S4.2.1     \\ 
$L$            & 5000 $L_\odot$ &  \S4.2.1     \\
$T$            & 3100 $K$         &  \S4.2.1   \\
$R$            & 280 $R_\odot$   &  \S4.2.1     \\
$M$            & 2 $M_\odot$     &  \S4.4       \\
 &          &     \\
 & White Dwarf:    &      \\
$L$            & 0.25 $L_\odot$ &  \S4.2.2     \\
$M$            &  $>$0.56 $M_\odot$ &  \S4.4   \\
 &          &     \\
 & Long Period Orbit: &   \\
$P$            &    15.6 $yr$  &  \S3          \\
$a$            &    8.5 $AU$   &  \S4.6        \\
$i$            &    84$^\circ$   &  \S4.6      \\
 &          &     \\
 & Circumstellar Shell:  &     \\
$R_{inner}$    & 22 $AU$          &  \S4.5     \\
\enddata
\end{deluxetable}

\clearpage

\begin{figure} 
\epsscale{0.8} 
\plotone{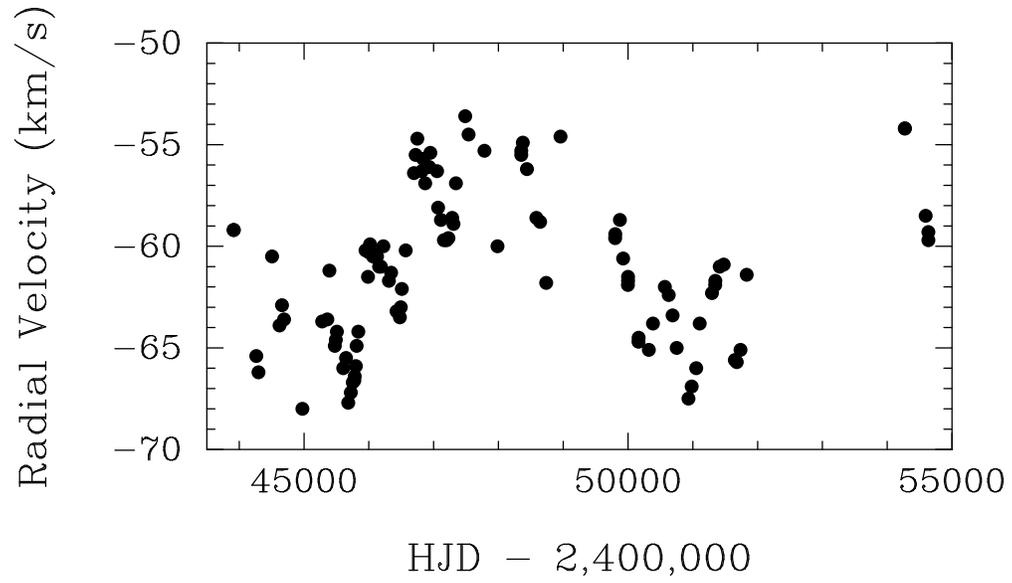}
\caption{Our 106 KPNO radial velocities, derived from lines in the
near infrared, of the CH~Cyg M giant component plotted versus 
heliocentric Julian day.  The data span nearly 30 years.  
A short period, 2.1 year, and a long period, 15.6 year, variation 
are apparent. 
}
\end{figure}

\clearpage

\begin{figure}
\epsscale{0.8}
\plotone{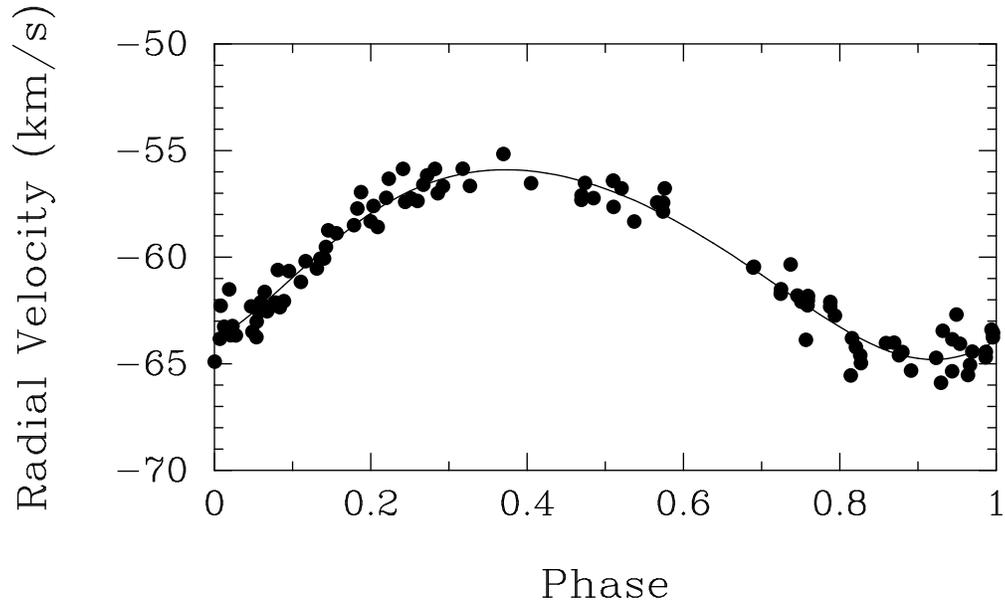}
\caption{The computed velocity curve of the 15.6 yr long period orbit 
compared with the KPNO radial velocities.  Each plotted velocity 
consists of the total observed velocity minus its calculated 
short period velocity.  Zero phase is a time of periastron. 
}
\end{figure}

\begin{figure}
\epsscale{0.8}
\plotone{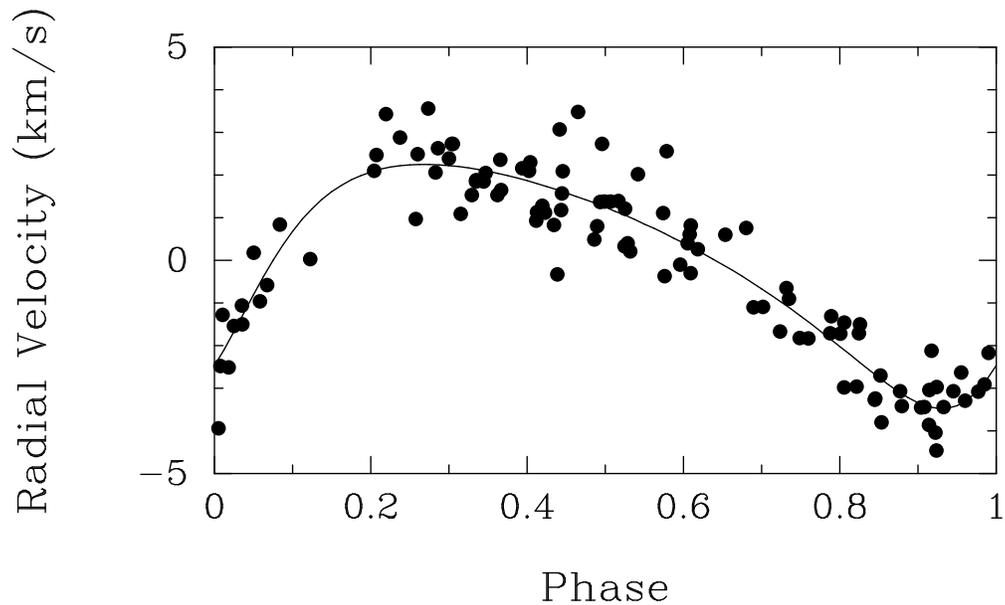}
\caption{The computed velocity curve of the 2.1 yr velocity variation
interpreted as an orbit, compared with the KPNO radial velocities.  
Each plotted velocity 
consists of the total observed velocity minus its calculated 
long period velocity.  Zero phase is a time of periastron.
}
\end{figure}

\clearpage

\begin{figure}
\epsscale{0.8}
\plotone{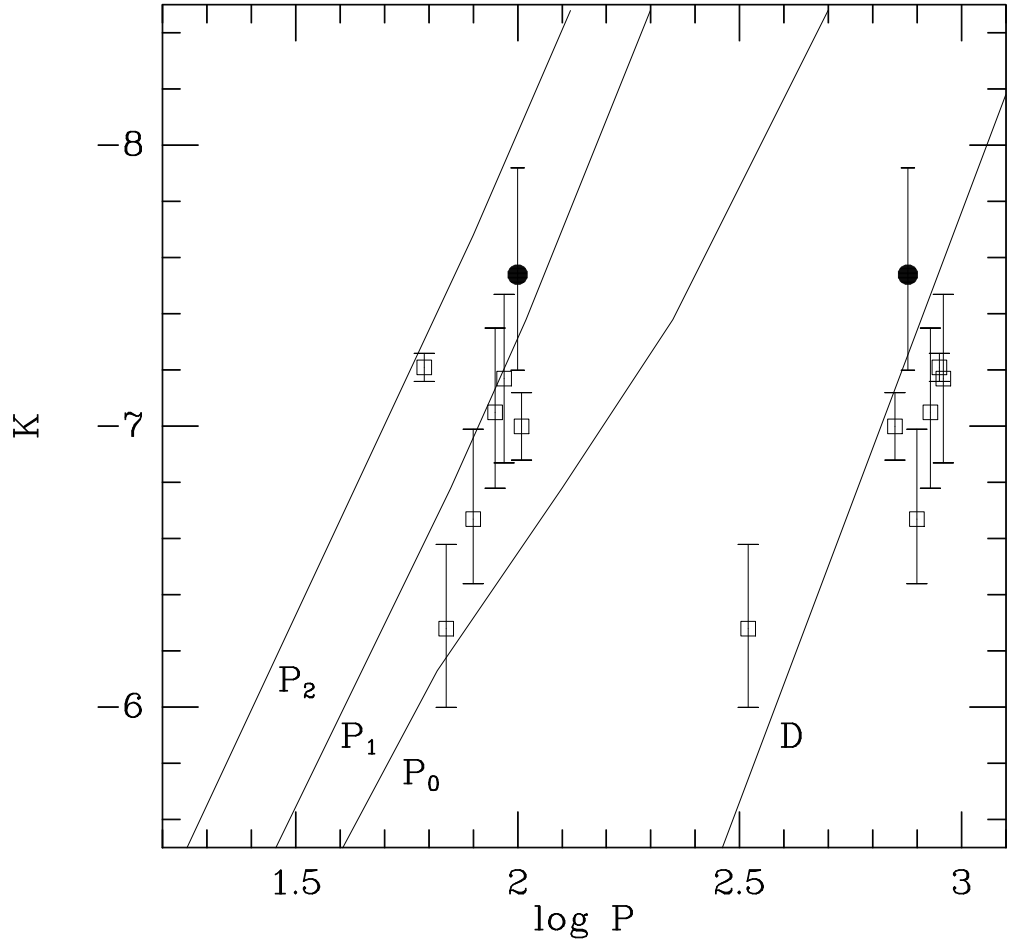}
\caption{Period--absolute $K$ magnitude plot of field LSP stars.  Each LSP star
has two periods, a normal pulsation mode and the LSP.  Filled circle
is CH~Cyg. Open squares are six LSP stars from \citet{hinkle_et_al_2002}
and \citet{wood_olivier_kawaler_2004} that have Hipparcos parallaxes.
The \citet{van_leeuwen_2007} revised Hipparcos parallax values are used. The
lines P$_0$, P$_1$, and P$_2$ show respectively linear fundamental, 
first overtone, and second overtone pulsation periods from
\citet{wood_et_al_1999} for oxygen-rich opacities.  The line D is
the middle range of the Wood's D relation \citep[see][]{wood_et_al_1999}.
For field stars the uncertainty in the distance, indicated by the
magnitude error bars, combined with the intrinsic range in magnitude
blurs the observed pulsation relations, although most of these stars
are clearly first overtone pulsators.
} 
\end{figure}

\begin{figure}
\epsscale{1.0}
\plottwo{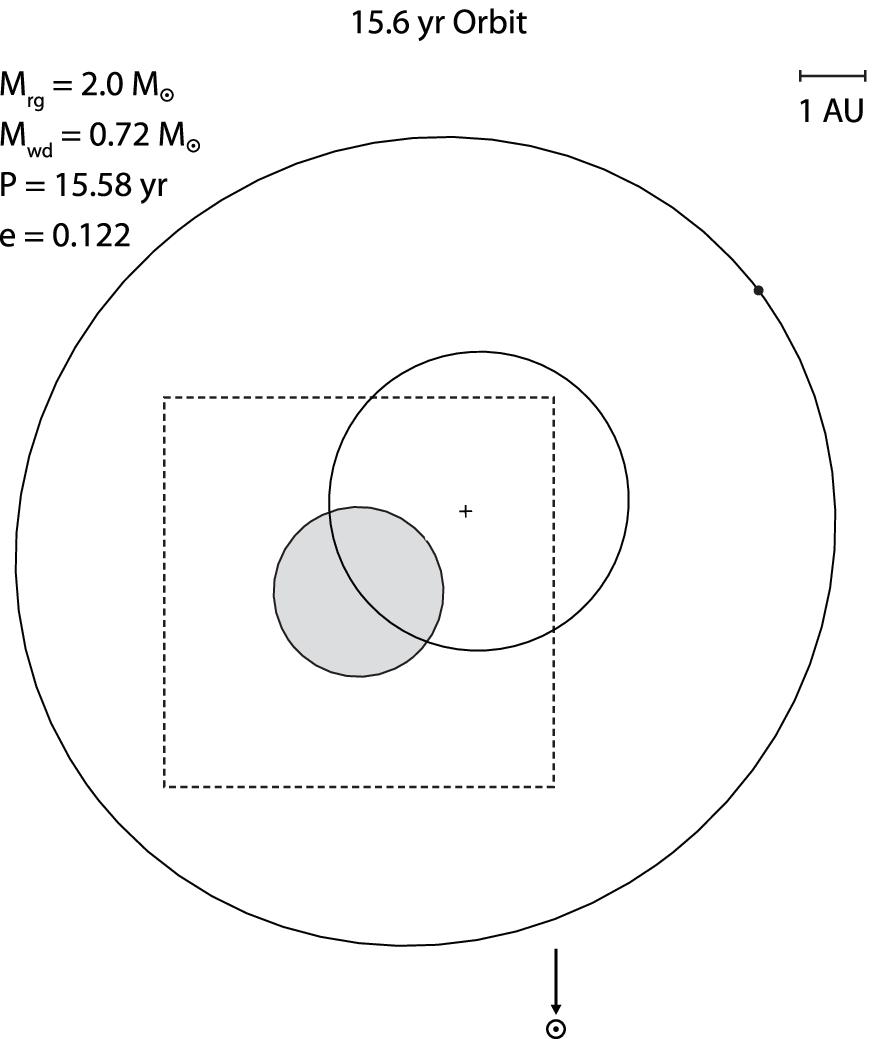}{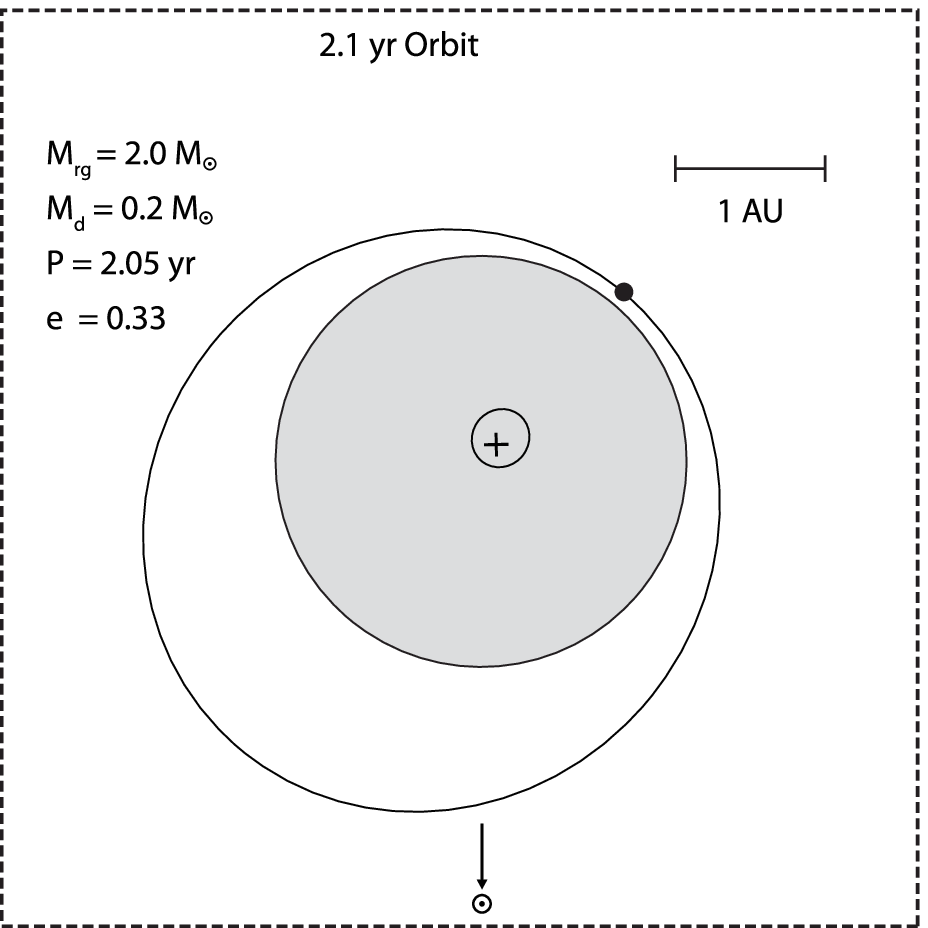}
\caption{
Polar view, to scale, of the computed orbits of the CH Cyg system.
On the left, the inner and outer ellipses are the 15.6 yr orbits with 
a red giant of mass 2 M$_\odot$ and a white dwarf 
of mass 0.72 M$_\odot$ (see text).  The
shaded circle and the dot represent the giant and white dwarf, respectively,
at periastron.
The shaded circle is the scaled size
(R=280 R$_\odot$) of the red giant which is assumed spherical.  
The plus is the center of mass of the orbit.
The dashed line box is shown enlarged on the right.  The inner and 
outer ellipses
are the 2.1 yr orbits of the red giant and the low mass companion shown at
periastron.  The plus sign is the center of mass of the 2.1 yr orbit.  
The 2.1 yr and 15.6 yr orbits are shown coplanar (see text).
The companion object in the 2.1 yr orbit can not be a white dwarf. 
However, we argue (see text) that the 2.1 year orbit is not physical.  
} 
\end{figure}

\begin{figure}
\epsscale{0.8}
\plotone{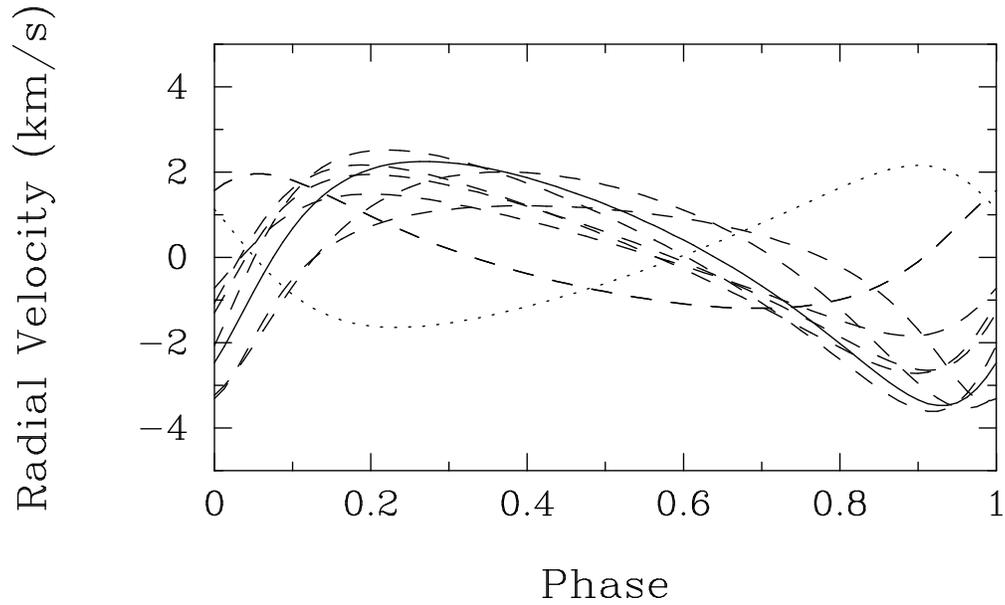}
\caption{A phase plot of the calculated radial velocity``orbits'' of 
long secondary periods for eight stars plus the 2.1 yr period of CH Cyg.  
Six stars, RS~CrB, AF~Cyg, X~Her, g~Her, V574~Oph and BI~Peg are from 
\citet{hinkle_et_al_2002}, and two, Z~Eri and S~Lep, are from 
\citet{wood_olivier_kawaler_2004}.  The center of mass velocity of 
each ``orbit'' has been subtracted from the observations to produce
velocity curves with the same zero point  for ease of comparison.
The ``orbit'' of CH~Cyg is a solid line, the other stars are 
dashed lines except for AF~Cyg, which is a dotted line to emphasize
that its ``orbit'' is in antiphase with the others.
}
\end{figure}


\begin{thebibliography}{}

\bibitem[Adams et al.(2006)]{adams_et_al_2006}
Adams, E., Wood, P. R., \& Cioni, M.-R. 2006, Mem. S.A.It., 77, 537

\bibitem[Baraffe \& Chabrier(1996)]{baraffe_chabrier_1996} 
Baraffe, I. \& Chabrier, G. 1996, \apj, 461, L51

\bibitem[Bessel \& Wood(1984)]{bessel_wood_84}
Bessel, M. S. \& Wood, P. R. 1984, \pasp, 96, 247

\bibitem[Biller et al.(2006)]{biller_et_al_2006}
Biller, B. A., Close, L. M., Li, A., Marengo, M., Bieging, J. H., 
Hinz, P. M., Hoffmann, W. F., Brusa, G., \& Miller, D. 2006, \apj, 647, 464

\bibitem[Crocker et al.(2002)]{crocker_et_al_2002}
Crocker, M. M., Davis, R. J., Spencer, R. E., Eyres, S. P. S., 
Bode, M. F., \& Skopal, A.  2002, \mnras, 335, 1100

\bibitem[Daniels(1966)]{d66}
Daniels, W. 1966, Univ. of Maryland, Dept. of Physics \& Astronomy
Technical Report No. 579

\bibitem[Derekas et al.(2006)]{derekas_et_al_2006}
Derekas, A., Kiss, L. L., Bedding, T. R., Kjeldsen, H., Lah, P.,
Szab{\'o}, Gy. M.  2006, \apjl, 650, L55

\bibitem[Deutsch et al.(1974)]{detal74}
Deutsch, A. H., Lowen, L., Morris, S. C., \& Wallerstein, G. 1974, \pasp,
86, 233

\bibitem[Dyck et al.(1998)]{dyck_et_al_1998}
Dyck, H. M., van Belle, G. T., \& Thompson, R. R. 1998, \aj, 116, 981

\bibitem[Eyres et al.(2002)]{eyres_et_al_2002}
Eyres, S. P. S., Bode, M. F., Skopal, A., Crocker, M. M., Davis, R. J.,
Taylor, A. R., Teodorani, M., Errico, L., Vittone, A. A., \& Elkin, V. G.
2002, \mnras, 335, 526

\bibitem[Ezuka et al.(1998)]{ezuka_et_al_1998}
Ezuka, H., Ishida, M., \& Makino, F. 1998, \apj, 499, 388

\bibitem[Feast et al.(1989)]{feast_et_al_1989}
Feast, M. W., Glass, I. S., Whitelock, P. A., \& Catchpole, R. M. 1989,
\mnras, 241, 375

\bibitem[Fekel(1997)]{fekel_1997}
Fekel, F. C. 1997, \pasp, 109, 514

\bibitem[Fekel et al.(2000)]{fetal00}
Fekel, F. C., Joyce, R. R., Hinkle, K. H., \& Skrutskie, M. F. 2000, \aj,
119, 1375

\bibitem[Fekel et al.(2003)]{fekel_et_al_03}
Fekel, F. C., Hinkle, K. H.,\& Joyce, R. R. 2003, in ASP Conf Ser.
303, ``Symbiotic Stars Probing Stellar Evolution,'' eds. R. L. M. Corradi,
J. Miko{\l}ajewska, T. J.  Mahoney (San Francisco:ASP), p. 113.

\bibitem[Fekel et al.(2007)]{fekel_et_al_2007}
Fekel, F. C., Hinkle, K. H., Joyce, R. R., Wood, P. R., \&
Lebzelter, T. 2007, \aj, 133, 17

\bibitem[Fekel et al.(2008)]{fekel_et_al_2008}
Fekel, F. C., Hinkle, K. H., Joyce, R. R., Wood, P. R., \&
Howarth, I. D. 2008, \aj, 136, 146

\bibitem[Fernie et al.(1986)]{fernie_et_al_1986}
Fernie, J.D., Lyons, R., Beattie, B., \& Garrison, R. F. 1986, IBVS, 2935

\bibitem[Fitzpatrick(1993)]{f93}
Fitzpatrick, M. J. 1993, in ASP Conf. Ser. 52, Astronomical Data Analysis
Software and Systems II, ed. R. J. Hanish, R. V. J. Brissenden, \&
J. Barnes (San Francisco:ASP), p. 472

\bibitem[Galloway et al.(2002)]{galloway_et_al_2002}
Galloway, D. K., Sokoloski, J. L., \& Kenyon, S. J. 
2002, \apj, 580, 1065

\bibitem[Gromadzki \& Miko{\l}ajewska(2008)]{gromadzki_mikolajewska_2008}
Gromadzki, M. \& Miko{\l}ajewska, J. 
2008, A\&A, submitted (astroph 0804.4139v1)

\bibitem[Hall et al.(1978)]{hetal78}
Hall, D. N. B., Ridgway, S. T., Bell, E. A., \& Yarborough, J. M. 1978,
Proc. SPIE, 172, 121

\bibitem[Hatzes(1996)]{hatzes_1996}
Hatzes, A. P. 1996, \pasp, 108, 839

\bibitem[Hatzes \& Cochran(1993)]{hatzes_cochran_1993}
Hatzes, A. P. \& Cochran, W. D. 1993, \apj, 413, 339

\bibitem[Hatzes \& Cochran(1999)]{hatzes_cochran_1999}
Hatzes, A. P \& Cochran, W. D. 1999, \mnras, 304, 109

\bibitem[Henry et al.(2000)]{henry_et_al_2000}
Henry, G. W., Fekel, F. C., Henry, W. M., \& Hall, D. S. 2000, \apjs, 130, 201

\bibitem[Hinkle et al.(1993)]{hinkle_et_al_93}
Hinkle, K. H., Fekel, F. C., Johnson, D. S., \& Scharlach, W. W.
G. 1993, \aj, 105, 1074

\bibitem[Hinkle et al.(1998)]{hetal98} 
Hinkle, K. H., Cuberly, R., Gaughan, N., Heynssens, J., Joyce, R.,
Ridgway, S., Schmitt, P., \& Simmons, J. E. 1998, Proc. SPIE, 3354,
810

\bibitem[Hinkle et al.(2002)]{hinkle_et_al_2002}
Hinkle, K. H., Lebzelter, T., Joyce, R. R., \& Fekel, F. C.  2002,
\aj, 123, 1002

\bibitem[Hinkle et al.(2006)]{hinkle_et_al_2006}
Hinkle, K. H., Fekel, F. C., Joyce, R. R., Wood, P. R., Smith, V., \&
Lebzelter, T. 2006, \aj, 641, 479

\bibitem[Hoard(1993)]{hoard_1993}
Hoard, D. W. 1993, \pasp, 105, 1232

\bibitem[Houk(1963)]{houk_1963}
Houk, N. 1963, \aj, 68, 253

\bibitem[Hughes \& Wood(1990)]{hughes_wood_1990}
Hughes, S. M. G. \& Wood, P. R. 1990, \aj, 99, 784

\bibitem[Hut(1981)]{hut_1981}
Hut, P. 1981, \aap, 99, 126

\bibitem[Iben \& Tutukov(1996)]{iben_tutukov_1996}
Iben, I., Jr. \& Tutukov, A. V. 1996, \apjs, 105, 145

\bibitem[Iben(2003)]{iben_2003}
Iben, I., Jr. 2003, in ASP Conference Sers.~303
``Symbiotic Stars Probing Stellar Evolution,''
eds.  R. L. M. Corradi, J. Miko{\l}ajewska, \& T. J. Mahoney 
(San Francisco:ASP), p. 177 

\bibitem[Joyce(1992)]{joyce_92} 
Joyce, R. R. 1992, in ASP Conf.  Ser.~23, ``Astronomical CCD Observing
and Reduction Techniques,'' ed.  S. Howell (San Francisco:ASP), p.
258

\bibitem[Joyce et al.(1998)]{jetal98}
Joyce, R. R., Hinkle, K. H., Meyer, M. R., \& Skrutskie, M. F. 1998, Proc.
SPIE, 3354, 741

\bibitem[Kalirai et al.(2008)]{kalirai_et_al_2007}
Kalirai, J. S., Hansen, B. M. S., Kelson, D. D., Reitzel, D. B.,
Rich, R. M., Richer, H. B. 2008, \apj, 676, 594

\bibitem[Karovska et al.(2007)]{karovska_et_al_2007}
Karovska, M., Carilli, C. L., Ramond, J. C., \& Mattei, J. A. 
2007, \apj, 661, 1048

\bibitem[Kenyon \& Fernandez-Castro(1987)]{kenyon_fernandez_1987}
Kenyon, S. J. \& Fernandez-Castro, T. 1987, \aj, 93, 938

\bibitem[Kenyon et al.(1988)]{kenyon_et_al_1988}
Kenyon, S. J., Fernandez-Castro, T., \& Stencel, R. E. 1988, \aj, 95, 1817

\bibitem[Kiss et al.(2000)]{kiss_et_al_2000}
Kiss, L. L., Szatm\'{a}ry, K., Szab\'{o}, G. \& Mattei, J. A. 2000, \aaps,
145, 283

\bibitem[Kotnik-Karuza et al.(2007)]{kotnik_et_al_2007}
Kotnik-Karuza, D., Jurkic, T., Friedjung, M. 2007, Baltic Astr., 16, 98

\bibitem[Lebzelter et al.(2000)]{lebzelter_et_al_2000}
Lebzelter, T., Kiss, L. L., \& Hinkle, K. H. 2000, \aap, 361, 167

\bibitem[Liebert et al.(2004)]{liebert_et_al_2004}
Liebert, J., Bergeron, P., Eisenstein, D., Harris, H. C., Kleinman, S. J., 
Nitta, A. \& Krzesinski, J. 2004, \apj, 606, L147

\bibitem[Lu \& Rucinski(1999)]{lu_rucinski_1999}
Lu, W. \& Rucinski, S. M. 1999, \aj, 118, 515

\bibitem[Luna \& Sokoloski(2007)]{luna_sokoloski_2007}
Luna, G. J. M. \& Sokoloski, J. L. 2007, \apj, 671, 741

\bibitem[Miko{\l}ajewska et al.(1988)]{mikolajewska_et_al_1988}
Miko{\l}ajewska, J., Selvelli, P.L., Hack, M. 1988, \aap, 198, 150

\bibitem[Miko{\l}ajewski et al.(1987)]{mikolajewski_et_al_1987}
Miko{\l}ajewski, M., Tomov, T., Miko{\l}ajewska, J.
1987, \apss, 131, 733

\bibitem[Miko{\l}ajewski et al.(1990a)]{mikolajewski_et_al_1990}
Miko{\l}ajewski, M., Miko{\l}ajewska, J., 
\& Khudiakova, T. N. 1990, \aap, 235, 219

\bibitem[Miko{\l}ajewski et al.(1990b)]{mikolajewski_et_al_1990b}
Miko{\l}ajewski, M., Miko{\l}ajewska, J., Tomov, T., Kulesza, B., 
Szczerba, R., \& Wikierski, B. 1990, Acta Astr., 40, 129

\bibitem[Miko{\l}ajewski et al.(1992)]{mikolajewski_et_al_1992}
Miko{\l}ajewski, M., Miko{\l}ajewska, \& Khudyakova, T. N. 1992,
\aap, 254, 127

\bibitem[Mikkola \& Tanikawa(1998)]{mikkola_tanikawa_1998}
Mikkola, S. \& Tanikawa, K. 1998, \aj, 116, 444

\bibitem[Muciek \& Miko{\l}ajewski(1989)]{muciek_mikolajewski_1989}
Muciek, M. \& Miko{\l}ajewski, M. 1989, Acta Astr., 39, 165

\bibitem[Munari et al.(1996)]{munari_et_al_1996}
Munari, U., Yudin, B. F., Kolotilov, E. A., \& Tomov, T. V. 
1996, \aap, 311, 484

\bibitem[M\"urset et al.(1991)]{murset_et_al_1991}
M\"urset, U., Nussbaumer, H., Schmid, H. M., \& Vogel, M. 
1991, \aap, 248, 458

\bibitem[M\"urset et al.(2000)]{murset_et_al_2000}
M\"urset, U., Dumm, T., Isenegger, S., Nussbaumer, H., Schild, H., Schmid,
H. M., \& Schmutz, W. 2000, \aap, 353, 952

\bibitem[Muterspaugh et al.(2008)]{muterspaugh_et_al_2008}
Muterspaugh, M. W., Lane, B. F., Fekel, F. C., Konacki, M., Burke, B. F., 
Kulkarni, S. R., Colavita, M. M., Shao, M., \& Wiktorowicz, S. J.
2008, \aj, 135, 766

\bibitem[Nussbaumer \& Vogel(1996)]{nussbaumer_vogel_1996}
Nussbaumer, H. \& Vogel, M. 1996, \aap, 307, 470

\bibitem[Olivier \& Wood(2003)]{olivier_wood_2003}
Olivier, E. A. \& Wood, P. R. 2003, \apj, 584, 1035

\bibitem[Payne-Gaposhkin(1954)]{payne-gaposhkin_1954}
Payne-Gaposhkin, C. 1954, Ann. Harvard College Observatory, 113, 189

\bibitem[Percy et al.(2004)]{percy_et_al_2004} 
Percy, J.R., Bakos, A. G., Besla, G., Hou, D., Velocci, V., Henry,
G. W. 2004, in ASP Conf. Sers.~310 ``Variable Stars in the Local
Group,'' ed.  D. W.  Kurtz \& Karen R. Pollard, (San Francisco:ASP),
p. 348

\bibitem[Podsiadlowski \& Mohamed(2007)]{podsiadlowski_mohamed_2007}
Podsiadlowski, Ph. \& Mohamed, S. 2007, in ``Evolution and Chemistry
of Symbiotic Star, Binary Post-AGB, and Related Objects''. J.
Miko{\l}ajewska \& R. Szczerba, eds., Baltic Astronomy, 16, 26

\bibitem[Prieur et al.(2002)]{prieur_et_al_2002}
Prieur, J. L., Aristidi, E., Lopez, B., Scardia, M., Mignard, F., \& 
Carbillet, M. 2002, \apjs, 139, 249

\bibitem[Richichi et al.(1999)]{richichi_et_al_1999} 
Richichi, A., Fabbroni, L., Ragland, S., \& Scholz, M. 1999, \aap, 344, 511

\bibitem[Rodgers et al.(1997)]{rodgers_et_al_1997}
Rodgers, B., Hoard, D. W., Burdullis, T., Machado-Pelaez, L., 
O'Toole, M., Reed, S.  1997, \pasp, 109, 1093

\bibitem[Russell et al.(1955)]{russell_et_al_1955}
Russell, H.N., Dugan, R. S., \& Stewart, J. Q. 1955, "Astronomy II --
Astrophysics \& Stellar Astronomy," (Ginn:Boston), p. 700

\bibitem[Scarfe et al.(1990)]{sbf90}
Scarfe, C. D., Batten, A. H., \& Fletcher, J. M. 1990, Publ. Dominion
Astrophys. Obs. Victoria, 18, 21

\bibitem[Schild et al.(1999)]{schild_et_al_1999}
Schild, H., Dumm, T., Folini, D., Nussbaumer, H., \& Schmutz, W. 1999, 
in ``The Universe as Seen by ISO,'' P. Cox \& M. F. Kessler, eds. (ESA-SP
427; ESA, Noordwijk), p. 397

\bibitem[Schmidt et al.(2006)]{schmidt_et_al_2006}
Schmidt, M. R., Za\v cs, L., Miko{\l}ajewska, J., \& Hinkle, K. H.
2006, \aap, 446, 603

\bibitem[Schmutz et al.(1994)]{schmutz_et_al_1994}
Schmutz, W., Schild, H., M\"urset, U., \& Schmid, H. M. 
1994, \aap, 288, 819

\bibitem[Skopal(1995)]{skopal_1995}
Skopal, A. 1995, IBVS, 4157

\bibitem[Skopal et al.(1996a)]{skopal_et_al_1996}
Skopal, A., Bode, M. F., Lloyd, H. M., Tamura, S. 1996a, \aap, 308, L9

\bibitem[Skopal et al.(1996b)]{skopal_et_al_1996b}
Skopal, A., Bode, M. F., Bryce, M., et al. 1996b, \mnras, 282, 327

\bibitem[Skopal(1997)]{skopal_1997}
Skopal, A. 1997, in ``Physical Processes in Symbiotic Binaries and Related 
Systems'' J. Miko{\l}ajewska, ed. (Copernicus Foundation for Polish Astronomy, 
Warsaw), p. 99

\bibitem[Skopal(1998)]{skopal_1998}
Skopal, A. 1998, \aap, 338, 599

\bibitem[Skopal et al.(2007)]{skopal_et_al_2007}
Skopal, A., Va\v nko, M., Pribulla, T., Chochol, D., Semkov, E., 
Wolf, M., \& Jones, A. 2007, Astron. Nachr., 328, 909

\bibitem[Soker(2000)]{soker_2000}
Soker, N. 2000, \aap, 357, 557

\bibitem[Sokoloski \& Kenyon(2003a)]{sokoloski_kenyon_2003a}
Sokoloski, J. L. \& Kenyon, S. J. 2003a, \apj, 584, 1021

\bibitem[Sokoloski \& Kenyon(2003b)]{sokoloski_kenyon_2003b}
Sokoloski, J. L. \& Kenyon, S. J. 2003b, \apj, 584, 1027

\bibitem[Solf(1987)]{solf_1987}
Solf, J. 1987, \aap, 180, 207

\bibitem[Soszy{\'n}ski(2007)]{soszynski_2007}
Soszy{\'n}ski, I. 2007, \apj, 660, 1486

\bibitem[Soszy{\'n}ski et al.(2004)]{soszynski_et_al_2004}
Soszy{\'n}ski, I., Udalski, A., Kubiak, M., Szyma{\'n}ski, M. K.,
Pietrzy{\'n}ski, G., {\.Z}ebru{\'n}, K., Szewczyk, O., Wyrzykowski,{\L}.,
Dziembowski, W. A., 2004, Acta, 54, 347

\bibitem[Soszy{\'n}ski et al.(2007)]{soszynski_et_al_2007}
Soszy{\'n}ski, I., Dziembowski, W. A., Udalski, A., Kubiak, M.,
Szyma{\'n}ski, M. K., Pietrzy{\'n}ski, G., Wyrzykowski, L., Szewczyk, O., \&
Ulaczyk, K. 2007, Acta, 57, 201

\bibitem[Szewczyk et al.(2008)]{szewczyk_et_al_2008}
Szewczyk, O., Pietrzy{\'n}ski, G., Gieren, W., Storm, J., Walker, A., 
Rizzi, L, Kinemuchi, K., Bresolin, F., Kudritzki, R.-P., \& Dall'Ora, M.
2008, \aj, 136, 272

\bibitem[Taranova \& Shenavrin(2004)]{taranova_shenavrin_2004}
Taranova, O. G. \& Shenavrin, V. I. 2004, Astronomy Reports, 48, 813

\bibitem[Taranova \& Shenavrin(2007)]{taranova_shenavrin_2007}
Taranova, O. G. \& Shenavrin, V. I. 2007, Astronomy Letters, 33, 531

\bibitem[Taylor et al.(1986)]{taylor_et_al_1986}
Taylor, A. R. , Seaquist, E. R., \& Mattei, J. A. 1986, Nature, 319, 38 

\bibitem[Unno et al.(1989)]{unno_et_al_1989}
Unno, W., Osaki, Y., Ando, H., Saio, H. \& Shibahashi, H. 1989, ``Nonradial
Oscillations of Stars,'' (Tokyo: Univ. of Tokyo Press), p. 32

\bibitem[van Leeuwen(2007)]{van_leeuwen_2007}
van Leeuwen, F. 2007, ``Hipparcos, the New Reduction of the Raw
Data,'' (Heidelberg :Springer)

\bibitem[Vassiliadis \& Wood(1993)]{vassiliadis_wood_93}
Vassiliadis, E. \& Wood, P. R. 1993, \apj 413, 641

\bibitem[Viotti et al.(1997)]{viotti_et_al_1997}
Viotti, R., Badiali, M., Cardini, D., Emanuele, A., Iijima, T.
1997, in ``Hipparcos Venice '97,'' B. Battrick, ed., ESA SP-402, p. 405

\bibitem[Walder et al.(2008)]{walder_et_al_2008}
Walder, R., Folini, D., \& Shore, S. N. 2008, \aap, 484, L9

\bibitem[Warner(1972)]{warner_1972}
Warner, B. 1972, \mnras, 159, 95

\bibitem[Webster \& Allen(1975)]{wa75}
Webster, B. L., \& Allen, D. A. 1975, \mnras, 171, 171

\bibitem[Whitelock(1987)]{whitelock_1987}
Whitelock, P. A. 1987, \pasp, 99, 573

\bibitem[Wolff(1996)]{wolff_1996}
Wolff, C.L. 1996, \apjl, 459, L103 

\bibitem[Wood(2007)]{wood_2007}
Wood, P. R. 2007, in ASP Conf. Series 362, ``7th Pacific Rim Conference
on Stellar Astrophysics,'' Y. W. Kang, H. W. Lee, K. S. Chen, \& K. C. Leung 
eds. (San Francisco:ASP), p. 234

\bibitem[Wood et al.(2004)]{wood_olivier_kawaler_2004}
Wood, P. R., Olivier, E. A., \& Kawaler, S. D. 2004, \apj, 604, 800

\bibitem[Wood et al.(1999)]{wood_et_al_1999}
Wood, P. R. et al. 1999, in ``Asymptotic Giant Branch Stars,'' IAU Symp. 191, 
T. Le Bertre, A. Lebre, C. Waelkens, eds., p. 151 

\bibitem[Wood \& Sebo(1996)]{wood_sebo_1996}
Wood, P. R. \& Sebo, K. M. 1996, \mnras, 282, 958

\bibitem[Wood \& Zarro(1981)]{wood_zarro_1981}
Wood, P. R. \& Zarro, D. M. 1981, \apj, 247, 247

\bibitem[Yamashita \& Maehara(1979)]{ym79}
Yamashita, Y., \& Maehara, H. 1979, PASJ, 31, 307

\bibitem[Yoo \& Yamashita(1991)]{yy91}
Yoo, K. H., \& Yamashita, Y. 1991, Publ. Natl. Astron. Obs. Japan, 2, 1

\bibitem[Zamanov et al.(2007)]{zamanov_et_al_2007}
Zamanov, R. K., Bode, M. F., Melo, C. H. F., Bachev, R., Gomboc,
A., Stateva, I. K., Porter, J. M., \& Pritchard, J. 2007, \mnras,
380, 1053

\end{thebibliography}
\end{document}